\documentclass[12pt]{amsart}
\usepackage{amsmath,amssymb,cite,enumerate,graphicx,hyperref,textcomp}

\hypersetup{pdftex,colorlinks=true,allcolors=blue}
\usepackage[all]{hypcap}

\usepackage[foot]{amsaddr}
\usepackage[sc]{mathpazo}

\newcommand\ack{\subsection*{Acknowledgment}}

\DeclareMathAlphabet\mathsfbi{T1}{phv}{b}{it}

\numberwithin{equation}{section}

\newcommand\BV{\boldsymbol} 
\newcommand\BM{\mathsfbi} 

\newcommand\dif{\:\!\mathrm{d}}
\newcommand\deriv[2]{\frac{\mathrm{d} #1}{\mathrm{d} #2}}
\newcommand\parderiv[2]{\frac{\partial #1}{\partial #2}}

\newcommand\RR{\mathbb R}

\newcommand\cL{\mathcal L}

\newcommand\cS{\mathcal S}
\newcommand\cB{\mathcal B}

\newcommand\diag{\mathrm{diag}}

\newcommand\hi{\hat i}
\newcommand\brho{\bar\rho}
\newcommand\bU{\bar{\BV U}}
\newcommand\bW{\bar{\BV W}}
\newcommand\bE{\,\bar{\!\BM E}}
\newcommand\bQ{\,\bar{\!\BM Q}}
\newcommand\bR{\,\bar{\!\BM R}}

\newcommand\myatop[2]{\genfrac{}{}{0pt}{}{#1}{#2}}
\newcommand\eqdef{\stackrel{\text{def}}{=}}

\begin{document}

\author[Rafail V. Abramov]{Rafail V. Abramov}

\address{Department of Mathematics, Statistics and Computer Science,
University of Illinois at Chicago, 851 S. Morgan st., Chicago, IL 60607}

\email{abramov@uic.edu}

\title{Turbulent energy spectrum via an interaction potential}

\begin{abstract}
For a large system of identical particles interacting by means of a
potential, we find that a strong large scale flow velocity can induce
motions in the inertial range via the potential coupling. This forcing
lies in special bundles in the Fourier space, which are formed by
pairs of particles. These bundles are not present in the Boltzmann,
Euler and Navier--Stokes equations, because they are destroyed by the
Bogoliubov--Born--Green--Kirkwood--Yvon formalism. However,
measurements of the flow can detect certain bulk effects shared across
these bundles, such as the power scaling of the kinetic energy.  We
estimate the scaling effects produced by two types of potentials: the
Thomas--Fermi interatomic potential (as well as its variations, such
as the Ziegler--Biersack--Littmark potential), and the electrostatic
potential.  In the near-viscous inertial range, our estimates yield
the inverse five-thirds power decay of the kinetic energy for both the
Thomas--Fermi and electrostatic potentials. The electrostatic
potential is also predicted to produce the inverse cubic power scaling
of the kinetic energy at large inertial scales. Standard laboratory
experiments confirm the scaling estimates for both the Thomas--Fermi
and electrostatic potentials at near-viscous scales. Surprisingly, the
observed kinetic energy spectrum in the Earth atmosphere at large
scales behaves as if induced by the electrostatic potential. Given
that the Earth atmosphere is not electrostatically neutral, we
cautiously suggest a hypothesis that the atmospheric kinetic energy
spectra in the inertial range are indeed driven by the large scale
flow via the electrostatic potential coupling.
\end{abstract}

\maketitle

\section{Introduction}

The phenomenon of turbulence in fluids has first been documented by
Leonardo da Vinci, and later by Boussinesq \cite{Bou} and Reynolds
\cite{Rey83,Rey}. As observed, turbulent motions in fluids appear to
be caused by the presence of a strong large scale flow, and manifest
in the ranges between the large and viscous scales (the so-called
``inertial range''). In 1941, Kolmogorov \cite{Kol41a,Kol41b,Kol41c}
suggested that the power scaling of the turbulent kinetic energy
spectrum could be modeled via an {\em ad hoc} dimensional
hypothesis. With help of Kolmogorov's hypothesis, Obukhov
\cite{Obu41,Obu49}, Chandrasekhar \cite{Cha49}, Corrsin \cite{Cor51},
and others observed that the time-averaged kinetic energy spectrum in
many real-world turbulent flows scales in the Fourier space as the
inverse five-thirds power of the wavenumber (known as the ``Kolmogorov
spectrum''). Although various other hypotheses on the nature of the
five-thirds spectrum have been proposed since then, none of those, to
our knowledge, offer a definitive physical explanation of why such a
spectrum manifests itself, relying instead upon additional {\em ad
  hoc} assumptions.

Curiously, the major proposed hypotheses for the turbulent kinetic
energy spectra have one feature in common -- namely, they rely upon
conventional models of fluid dynamics, such as the Euler or
Navier--Stokes equations. In the current work, we consider a
possibility that the main reason for this long standing difficulty
with the description of turbulence is that the standard fluid dynamics
equations lack the necessary physical effects to naturally recover
turbulent features in a flow. In order to see why this might happen,
here we start with reviewing the microscopic dynamics of fluid motion
and the standard derivation of the Navier--Stokes and Euler equations
from the kinetic model of a multiparticle fluid, and point out
possible reasons for such a peculiar deficiency.

Within the scope of the classical mechanics, a fluid consists of many
identical particles, each pair of which interact (mostly repel, but
sometimes attract) via a potential. If the potential function is
known, then one can explicitly formulate the system of first-order
ordinary differential equations (ODE) for the coordinate and velocity
of each particle as functions of time. Using the vector field of this
system, one can also derive the first-order linear partial
differential equation (PDE) for the probability density of states of
this system, which is known as the Liouville
\cite{Cer,CerIllPul,GalRayTex}, forward Kolmogorov
\cite{Abr17,GikSko,Oks}, or Fokker--Planck \cite{Ris} equation.

For a system of $N$ particles, the Liouville equation is a
$3N$-dimensional PDE, and, of course, cannot be solved explicitly in
real-world scenarios (where $N\sim 10^{23}$). For a practical
computation, the following simplifications are made. First, it is
assumed that the interaction potential has a very short range, so that
its effect can be approximated via a ``hard sphere collision''
interaction \cite{Bol,Gra,Cer,CerIllPul}. Second, it is assumed that
the probability density of the complete system is symmetric under the
permutations of particles (that is, the particles behave statistically
identically). This allows to use the
Bogoliubov--Born--Green--Kirkwood--Yvon (BBGKY) formalism
\cite{Bog,BorGre,Kir} to obtain a PDE for the marginal probability
density of a single particle, which, however, depends on the joint
two-particle distribution. Third, the additional assumption of
statistical independence of all particles is used to express the
two-particle probability distribution as a product of two
single-particle distributions. The resulting standalone equation for
the probability distribution of a single particle is known as the
Boltzmann equation \cite{Bol,Cer,CerIllPul}. The effects of
interactions between different particles are approximated by the
collision integral of the Boltzmann equation.

To obtain the Euler or Navier--Stokes equations, the Boltzmann
equation is further integrated against various powers of the velocity
variable. The zero-order moment equation becomes the evolution
equation for the density of the fluid, the first-order moment equation
describes the velocity, while the contracted second-order moment
equation is the one for the energy. Due to the mass, momentum and
energy conservation laws, the corresponding velocity moments of the
Boltzmann collision integral disappear from these three moment
equations.  Further, if the higher-order non-Gaussian moments of the
solution are presumed to be zero, the compressible Euler equations
emerge \cite{Gols}. If however, the higher-order moments are
parameterized from the equations for the stress and heat flux via the
Chapman--Enskog expansion \cite{ChaCow,HirCurBir} and the resulting
Newton and Fourier laws of viscosity and heat conductivity,
respectively, then the compressible Navier--Stokes equations are
obtained \cite{Gra}. In many practical applications, the flows are
effectively incompressible, in which case the zero-order density
moment is treated as a constant, leading to the incompressible Euler
and Navier--Stokes equations.

As we can see, in order to arrive at the standard equations of fluid
dynamics from the Liouville equation, one has to make some drastic
simplifications, which are based on rather strong assumptions. In the
present work, we entertain a hypothesis that the apparent lack of
turbulent effects in conventional models of fluid mechanics is caused
by these simplifications. To this end, we directly investigate the
Liouville equation for $N$ particles, which interact via a generic
potential.  What we uncover is that a strong large scale flow velocity
creates forcing in the inertial scales via the potential. This forcing
manifests in special $3$-dimensional bundles of the full
$3N$-dimensional coordinate space, which are shared by all unordered
pairs of particles. As suspected, these bundles are indeed destroyed
in the course of the BBGKY formalism, so that no such forcing
manifests in the Boltzmann equation, and, subsequently, in the Euler
and Navier--Stokes equations.  At the same time, it turns out that
measurements can detect some bulk effects of the strong large scale
flow forcing, if they are shared across these bundles.

Next, we estimate the scaling of suitably windowed time averages of
solutions of the Liouville equation for $N$ particles, which interact
via either the Thomas--Fermi interatomic potential \cite{Tho,Fer} (and
variations, such as the Ziegler--Biersack--Littmark potential
\cite{ZieBieLit}), or the electrostatic potential. Assuming that the
inertial range motions are driven by a strong large scale flow
velocity via a potential, for a suitable averaging time scale we find
that:
\begin{enumerate}[\indent a.]
\item In a small scale turbulent flow (in the proximity of the viscous
  range), the time averages of the kinetic energy should scale as the
  inverse five-thirds power of the wavenumber for both the
  Thomas--Fermi and the electrostatic potentials;
\item In a larger scale turbulent flow, the time averages of the
  kinetic energy should scale as the inverse cubic power of the
  wavenumber for the electrostatic potential. On the other hand, the
  Thomas--Fermi potential has a short effective range, and thus is not
  expected to affect the motion at large scales in any specific way.
\end{enumerate}
An interesting observation is that the physical mechanism of creation
of these energy spectra is absent from the conventional Euler and
Navier--Stokes equations, since the latter do not contain the
potential interaction terms in the energy equation.

To compare the theoretically predicted turbulent energy spectra with
the observations, we refer to the works of Buchhave and Velte
\cite{BucVel}, and Nastrom and Gage \cite{NasGag}. In \cite{BucVel},
the turbulent energy spectra are measured in laboratory conditions,
and exhibit the scaling behavior consistent with the predictions for
the Thomas--Fermi potential. In \cite{NasGag}, the turbulent energy
spectra are computed from direct observations of the atmospheric flow.
Surprisingly, the decay of the turbulent energy spectra in
\cite{NasGag} is more consistent with the electrostatic potential,
since they have the evidence of both the inverse five-thirds power
scaling at small scales, and the inverse cubic power scaling at larger
scales. Given that the Earth atmosphere is not electrostatically
neutral, we suggest a hypothesis that the atmospheric turbulent energy
spectra could indeed be produced by the large scale flow coupled to
the inertial ranges via the electrostatic potential.

\section{A system of interacting particles and its Liouville equation}

We start with a system of $N$ identical particles, which interact via
a potential $\phi(r)$. Here, any given particle, situated at the
coordinate point $\BV y$, creates the potential field $\phi(\|\BV
x-\BV y\|)$ around itself. Subsequently, any other particle, placed at
the point $\BV x$, experiences the acceleration given via
\begin{equation}
\BV a=-\parderiv{}{\BV x}\phi(\|\BV x-\BV y\|).
\end{equation}
The total acceleration, experienced by an $i$-th particle, is thus given
as the sum of contributions from the remaining particles:
\begin{equation}
\BV a_i=-\sum_{\myatop{j=1}{j \neq i}}^N\parderiv{}{\BV
  x_i}\phi\left(\|\BV x_i-\BV x_j\|\right).
\end{equation}
Above, $\BV x_i$ and $\BV x_j$ are the coordinate of $i$-th and $j$-th
particles, respectively.

Knowing the accelerations, we can construct the system of equations of
motion for the complete system of $N$ particles as
\begin{equation}
\label{eq:dyn_sys}
\deriv{\BV x_i}t=\BV v_i,\qquad\deriv{\BV v_i}t=\BV a_i=
-\sum_{\myatop{j=1}{j \neq i}}^N\parderiv{}{\BV x_i}\phi\left(\|\BV
x_i-\BV x_j\|\right),
\end{equation}
where $\BV v_i$ is the velocity of the $i$-th particle. For further
convenience, we concatenate $\BV X=(\BV x_1,\ldots,\BV x_N)$, $\BV
V=(\BV v_1,\ldots,\BV v_N)$, and denote the combined potential via
$\Phi(\BV X)$:
\begin{equation}
\label{eq:Phi}
\Phi(\BV X)=\sum_{i=1}^{N-1}\sum_{j=i+1}^N\phi\left(\|\BV x_i-\BV x_j
\|\right).
\end{equation}
Then, the system of equations of motion in \eqref{eq:dyn_sys} can be
written in a vector form via
\begin{equation}
\deriv{}t\left(\begin{array}{c}\BV X\\\BV V\end{array}\right)=
  \left(\begin{array}{c}\BV V\\-\partial\Phi/\partial\BV
    X \end{array}\right).
\end{equation}
Let $F(t,\BV X,\BV V)$ be the probability distribution for the whole
system of $N$ particles, in the sense that, at time $t$, $F(t,\BV
X,\BV V)\dif\BV X\dif\BV V$ is the probability that the system can be
found in the elementary phase volume $\dif\BV X\dif\BV V$ adjacent to
the state $(\BV X,\BV V)$. Then, the corresponding Liouville
\cite{Cer,CerIllPul,GalRayTex} (also known as the forward Kolmogorov
\cite{Abr17,GikSko,Oks} or Fokker--Planck \cite{Ris}) equation for $F$
is given via
\begin{equation}
\label{eq:liouville}
\parderiv Ft+\BV V\cdot\parderiv F{\BV X}=\parderiv\Phi{\BV X}\cdot
\parderiv F{\BV V}.
\end{equation}
In what follows, it is convenient to transform the Liouville equation
in a manner that removes the high-dimensional differentiation
operators.  We will use two different techniques for eliminating each
of the operators -- the moment integration for $\BV V$, and the
Fourier transformation for $\BV X$.

\subsection{The equations for velocity moments}

To eliminate the $\BV V$-derivative in the right-hand side of the
Liouville equation \eqref{eq:liouville}, we integrate both sides of
\eqref{eq:liouville} against various powers of $\BV V$, such that the
$\BV V$-derivative in the right-hand side of \eqref{eq:liouville} can
be removed via the integration by parts. This is a standard approach
for the Boltzmann equation \cite{Bol,Cer}, which leads to the
conventional Euler equations \cite{Gols}. We use the following
standard notations for the velocity moments: the density $\rho$,
average velocity vector $\BV U$, average kinetic energy matrix $\BM
E$, and the cubic moment tensor $\BM M_3$:
\begin{subequations}
\begin{equation}
\rho=\int_{\RR^{3N}} F\dif\BV V,\qquad\rho\BV U=\int_{\RR^{3N}}\BV V
F\dif\BV V,
\end{equation}
\begin{equation}
\rho\BM E=\int_{\RR^{3N}}\BV V^2 F\dif\BV V, \qquad \rho\BM M_3=
\int_{\RR^{3N}}\BV V^3 F\dif \BV V,
\end{equation}
\end{subequations}
where by ``$\BV V^k$'' we denote the outer product of $k$
vectors. Integrating \eqref{eq:liouville} against the outer powers of
$\BV V$ above, and assuming that no surface integrals emerge, we
arrive at
\begin{subequations}
\label{eq:moments_liouville_2}
\begin{equation}
\parderiv\rho t+\parderiv{}{\BV X}\cdot(\rho\BV U)=0, \qquad
\parderiv{(\rho\BV U)}t+\parderiv{}{\BV X}\cdot(\rho\BM E)=
-\rho\parderiv\Phi{\BV X},
\end{equation}
\begin{equation}
\parderiv{(\rho\BM E)}t+\parderiv{}{\BV X}\cdot(\rho\BM M_3)=
-\rho\bigg(\parderiv\Phi{\BV X}\BV U^T+\BV U\parderiv\Phi{\BV X}^T
\bigg).
\end{equation}
\end{subequations}
In what is to follow, it is convenient to express the cubic moment
$\BM M_3$ via the corresponding centered moment (or {\em skewness}).
Namely, let us write the identity
\begin{multline}
\BV V^3=(\BV V-\BV U)^3+\BV U\otimes\BV V\otimes\BV V+\BV V\otimes\BV
U\otimes\BV V+\BV V\otimes\BV V\otimes\BV U-\\-\BV V\otimes\BV U
\otimes\BV U- \BV U\otimes\BV V\otimes\BV U-\BV U\otimes\BV U\otimes
\BV V+\BV U^3,
\end{multline}
where ``$\otimes$'' denotes the outer product. Then, denote the
skewness tensor $\BM Q$ via the centered cubic moment
\begin{equation}
\label{eq:skewness}
\rho\BM Q=\int_{\RR^{3N}}(\BV V-\BV U)^3 F\dif\BV V.
\end{equation}
This allows to express the uncentered cubic moment $\BM M_3$ via
\begin{subequations}
\begin{equation}
\rho\BM M_3=\int_{\RR^{3N}}\BV V^3 F\dif\BV V=\rho(\BM Q+\BM R),
\end{equation}
\begin{equation}
\BM R=\BV U \otimes\BM E+(\BV U\otimes\BM E)^T+(\BV U\otimes\BM
E)^{TT}-2\BV U^3,
\end{equation}
\end{subequations}
where ``$T$'' and ``$TT$'' denote the two cyclic permutations of the
indices of a rank 3 tensor (which can be interpreted as two
``transpositions'', analogously to the single transposition of a
matrix).
The moment equations in \eqref{eq:moments_liouville_2} become
\begin{subequations}
\label{eq:moments_liouville}
\begin{equation}
\parderiv\rho t+\parderiv{}{\BV X}\cdot(\rho\BV U)=0, \qquad
\parderiv{(\rho\BV U)}t+\parderiv{}{\BV X}\cdot(\rho\BM E)=
-\rho\parderiv\Phi{\BV X},
\end{equation}
\begin{equation}
\parderiv{(\rho\BM E)}t+\parderiv{}{\BV X}\cdot(\rho(\BM Q+\BM R))
=-\rho\bigg(\parderiv\Phi{\BV X}\BV U^T+\BV U\parderiv\Phi{\BV X}^T
\bigg).
\end{equation}
\end{subequations}

\subsection{The Fourier transformation of the moment equations}

To remove the $\BV X$-derivative, the suitable technique is the
Fourier transformation, since it replaces the differentiation
operators with the wavevector multiplications. In what follows, we
adopt the notation
\begin{equation}
\hi=2\pi\sqrt{-1},
\end{equation}
to be able to use the letter ``$i$'' for indexing the summations
below.  Applying the Fourier transformation in $\BV X$ to
\eqref{eq:moments_liouville}, and denoting the wavevector as $\BV
K=(\BV k_1,\ldots,\BV k_N)$, we arrive at
\begin{subequations}
\label{eq:moments_fourier}
\begin{equation}
\frac 1\hi\deriv{\rho_{\BV K}}t+\BV K\cdot(\rho_{\BV K}\ast\BV U_{\BV
  K})=0,
\qquad
\frac 1\hi\deriv{}t(\rho_{\BV K}\ast\BV U_{\BV K})+\BV K \cdot(
\rho_{\BV K}\ast\BM E_{\BV K})=-\rho_{\BV K}\ast(\BV K\Phi_{\BV K}),
\end{equation}
\begin{equation}
\frac 1\hi\deriv{}t(\rho_{\BV K}\ast\BM E_{\BV K})+\BV K\cdot (\rho_{
  \BV K}\ast(\BM Q_{\BV K}+\BM R_{\BV K}))=-\rho_{\BV K}\ast \left[
  (\BV K\Phi_{\BV K})\ast\BV U_{\BV K}^T+\BV U_{\BV K}\ast(\BV K
  \Phi_{\BV K})^T\right],
\end{equation}
\end{subequations}
where ``$\ast$'' denotes the convolution in the Fourier space. The
Fourier transform $\BM R_{\BV K}$ is given via
\begin{equation}
\label{eq:RK}
\BM R_{\BV K}=\BV U_{\BV K}\circledast\BM E_{\BV K} +(\BV U_{\BV K}
\circledast\BM E_{\BV K})^T+(\BV U_{\BV K} \circledast\BM E_{\BV K}
)^{TT}-2\BV U_{\BV K}\circledast\BV U_{\BV K}\circledast\BV U_{\BV K},
\end{equation}
where ``$\circledast$'' denotes the superposition of a convolution
with an outer product.

\section{Potential-induced forcing in the inertial range}
\label{sec:structures}

What we obtained in \eqref{eq:moments_fourier} is an
infinite-dimensional system of nonlinear ODE for the Fourier
transforms $\rho_{\BV K}$, $\BV U_{\BV K}$ and $\BM E_{\BV K}$, with
$\BM Q_{\BV K}$ being a set of ``free parameters''. Obviously, this
system has a broad range of formal solutions, and only a small subset
of those is physically relevant. First, observe that, in the context
of physics, the Fourier transform $\BM Q_{\BV K}$ of the skewness
tensor in \eqref{eq:skewness} is not a free parameter, but rather
obeys its own set of equations, which involve even higher-order
moments, and so forth (see the moment closure problem of molecular
kinetics \cite{ChaCow,HirCurBir,Lev}).  Second, even if we are somehow
able to take all of those higher-order moments into account -- which,
effectively, amounts to solving the averaged Liouville equation in
\eqref{eq:liouville} -- the latter will still have many solutions
which are not physically relevant; for example, those where the
particles are arranged in way so as to have a low-dimensional periodic
motion, or even to avoid any interaction at all.  Thus, in order to
make meaningful conclusions from \eqref{eq:moments_fourier}, we need
to place qualitative restrictions on its solutions of interest, which
correspond to a realistic behavior of the observed flow.

In practical scenarios, the typical observed structure of the system
(say, the flow of a gas) consists of three distinct ranges of spatial
scales:
\begin{enumerate}[\indent a.]
\item {\bf The large scale flow.} Typically, there is a relatively
  small cluster of low-order Fourier wavenumbers, which contain a
  strong self-coupled solution describing the observed macroscopic
  phenomenon (for example, a jet or an eddy). This strong large-scale
  flow is typically ``self-coupled'', in the sense that it can be
  accurately described by a reduced, simplified system which only
  couples together these large scale motions and excludes small-scale
  effects (for example, the stratified rotating Boussinesq,
  barotropic, or quasigeostrophic equations for Earth's atmosphere).
  In what follows, we denote the Fourier transform of this strong
  large scale flow via $\BV W_{\BV K}$ (for convenience, we use a
  different letter for the strong large scale flow velocity to
  differentiate it from the ``generic'' inertial range velocity $\BV
  U_{\BV K}$).
\item {\bf The viscous range.} At the opposite side of the spectrum,
  on the scales of the average distance between the particles, the
  effect of the interaction potential $\phi$ becomes strong. As a
  result, the Fourier transform of the flow at small scales becomes
  nonlinearly coupled to itself via the interaction potential. The
  statistical macroscopic effect of this self-coupling is known as the
  {\em viscosity}, and can be derived in a standard fashion from the
  Boltzmann equation via the Chapman--Enskog expansion
  \cite{Abr17,ChaCow,HirCurBir,Gra}. The result is that the Fourier
  transform of the velocity decays exponentially rapidly in time at
  small scales, constituting an effective ``cut-off'' of the average
  velocity in the viscous range. To balance out the equations, this,
  in turn, triggers the growth of higher-order moments such as the
  skewness, the effects of which are observed in the form of the heat
  fluxes and the Fourier law of heat conductivity.
\item {\bf The inertial range.} The remaining range of the Fourier
  wavenumbers lies between the large scale flow and the viscous
  scales, and is known as the {\em inertial range}. On these scales,
  the effect of the interatomic potential $\phi$ is not strong enough
  to cause the effects of self-coupling, but, at the same time, it is
  still strong enough to induce motions via the coupling to the strong
  large-scale flow velocity $\BV W_{\BV K}$. This is the Fourier
  wavenumber range where the turbulent effects are observed.
\end{enumerate}
In what follows, the wavenumber $\BV K$ in the moment equations
\eqref{eq:moments_fourier} refers to the inertial range. In this
range, we assume that the effect of the potential forcing can be
completely neglected, with the exception of terms which are coupled to
the large scale flow velocity $\BV W_{\BV K}$. This leads to the
following simplification of \eqref{eq:moments_fourier} in the inertial
range:
\begin{subequations}
\label{eq:moments_fourier_simplified}
\begin{equation}
\frac 1\hi\deriv{\rho_{\BV K}}t+\BV K\cdot(\rho_{\BV K}\ast\BV U_{\BV
  K})=0,\qquad\frac 1\hi\deriv{}t(\rho_{\BV K}\ast\BV U_{\BV K})+\BV K
\cdot(\rho_{\BV K}\ast\BM E_{\BV K})=0,
\end{equation}
\begin{equation}
\frac 1\hi\deriv{}t(\rho_{\BV K}\ast\BM E_{\BV K})+\BV K\cdot
(\rho_{\BV K}\ast(\BM Q_{\BV K}+\BM R_{\BV K}))=-\rho_{\BV K}\ast
\left[ (\BV K\Phi_{\BV K})\ast\BV W_{\BV K}^T+\BV W_{\BV K}\ast(\BV K
  \Phi_{\BV K})^T\right].
\end{equation}
\end{subequations}
Here, the Fourier transform $\BV W_{\BV K}$ of the strong large scale
flow velocity is treated as an ``external forcing'', as we assume that
the motions in the inertial range are weak enough to not cause any
substantial feedback to the large scales via the potential. The
corresponding damping is ``hidden'' in $\BM Q_{\BV K}$, and manifests
as the Fourier law of heat conductivity at appropriate spatial and
temporal scales.

Let us further assume that, at the initial time, the Fourier
transforms $\rho_{\BV K}$, $\BV U_{\BV K}$, $\BM E_{\BV K}$ and $\BM
Q_{\BV K}$ in the inertial range are zero (that is, the flow is not
turbulent initially). This means that any motions in the inertial
range must be caused by the strong large scale flow velocity $\BV
W_{\BV K}$ via the potential coupling in the energy equation in
\eqref{eq:moments_fourier_simplified}. Thus, in order to understand
how turbulent motions are produced in the inertial range, we need to
examine the structure of the forcing induced via the strong large
scale flow velocity $\BV W_{\BV K}$.

For the sake of simplicity of the argument, let us assume that the
solution of \eqref{eq:moments_fourier_simplified} is nearly
incompressible (which is indeed the case in many observed turbulent
flows), and thus the Fourier transform $\rho_{\BV K}$ is largely
restricted to its zero Fourier coefficient $\rho_{\BV 0}$. For the
energy equation, it means that the time derivative of the energy $\BM
E_{\BV K}$, as well as the advection term $\BV K\cdot(\BM Q_{\BV
  K}+\BM R_{\BV K})$, are affected directly by the sum of the
convolution $(\BV K\Phi_{\BV K})\ast\BV W_{\BV K}^T$ with its own
transpose. In order to examine the structure of this convolution, let
us write it directly via the integral:
\begin{equation}
\label{eq:KPhiW}
(\BV K\Phi_{\BV K})\ast\BV W_{\BV K}^T=\int_{\RR^{3N}}(\BV K-\BV K')
  \Phi_{\BV K-\BV K'}\BV W_{\BV K'}^T\dif\BV K'.
\end{equation}
To proceed, we need to know the structure of the Fourier transform
$\Phi_{\BV K}$ of the combined potential.  In Appendix
\ref{app:potential_structure}, we show in a straightforward fashion
that $\Phi_{\BV K}$ has the form
\begin{equation}
\label{eq:PhiK}
\Phi_{\BV K}=\sum_{i=1}^{N-1}\sum_{j=i+1}^N\phi_{\|\BV k_i\|} \delta(
\BV k_i+\BV k_j)\prod_{\myatop{m=1}{m\neq i,j}}^N\delta(\BV k_m),
\end{equation}
with $\phi_{\|\BV k\|}$ given via
\begin{equation}
\label{eq:phik}
\phi_{\|\BV k\|}=\frac 2{\|\BV k\|}\int_0^\infty r\phi(r)
\sin(2\pi\|\BV k\|r)\dif r.
\end{equation}
Observe that the Fourier transform $\Phi_{\BV K}$ in \eqref{eq:PhiK}
is zero for almost all $\BV K\in\RR^{3N}$; in order for $\Phi_{\BV K}$
to be nonzero, $\BV K$ must belong to a $3$-dimensional subspace
$\cS_{ij}$ of the form
\begin{equation}
\label{eq:S_ij}
\cS_{ij}=\{\BV K=(\BV 0,\ldots,\BV 0,\stackrel{i\text{th}}{\BV k},\BV
0,\ldots,\BV 0,\stackrel{j\text{th}}{-\BV k},\BV 0,\ldots,\BV
0),\;\forall\BV k\in\RR^3\}.
\end{equation}
In addition, if $\BV K\in\cS_{ij}$, then only one term in the
summation in \eqref{eq:PhiK} is nonzero, which is when the summation
indices match those of the subspace $\cS_{ij}$ (unless $\BV k=\BV 0$,
in which case all terms in the summation are identical). For a system
consisting of $N$ particles, there are $N(N-1)/2$ such spaces
$\cS_{ij}$ in total, each corresponding to an unordered pair of
distinct particles.

Now that we have examined the structure of $\Phi_{\BV K}$, we can look
at the structure of the forcing $(\BV K\Phi_{\BV K})\ast\BV W_{\BV
  K}^T$.  If we compare \eqref{eq:KPhiW} with \eqref{eq:PhiK} and
\eqref{eq:S_ij}, it becomes clear that the nonzero values of the
integrand in \eqref{eq:KPhiW} are achieved for $\BV K-\BV
K'\in\cS_{ij}$. At the same time, $\BV W_{\BV K'}$ is the Fourier
transform of the large scale flow velocity, and, therefore, is nonzero
only for a set of small $\BV K'$, say, $\|\BV K'\|<b_{\BV W}\ll\|\BV
K\|$ (where $b_{\BV W}$ can be viewed as the ``spectral bandwidth'' of
the strong large scale flow velocity $\BV W$). We conclude that the
nonzero values of \eqref{eq:KPhiW} lie in a {\em bundle}
\begin{equation}
\label{eq:B_ij}
\cB_{ij}=\{\BV K-\BV K'\in\cS_{ij},\;\forall\|\BV K'\|<b_{\BV W}\}.
\end{equation}
It is clear that the bundle $\cB_{ij}$ is simply a relatively ``thin''
collection of translations of $\cS_{ij}$ around the origin to the
distance of at most $b_{\BV W}$. Also, any two distinct bundles
$\cB_{ij}$ only intersect in a small region around zero (which
corresponds to large scales), and are disjoint in the inertial
range. Within each $\cB_{ij}$, $\BV K$ is roughly of the form
\eqref{eq:S_ij}, being somewhat ``detuned'' by at most $b_{\BV W}$.

Lastly, let us assume that the Fourier transform $\phi_{\|\BV k\|}$ in
\eqref{eq:phik} scales as some power of $\|\BV k\|$ in the inertial
range. Then, it is obvious that, first, in $\Phi_{\BV K}$, given via
\eqref{eq:PhiK}, the same power scaling is shared across all subspaces
$\cS_{ij}$, defined by \eqref{eq:S_ij}; and, second, that in the
forcing \eqref{eq:KPhiW}, the same power scaling in the inertial range
is shared across all bundles $\cB_{ij}$, defined via \eqref{eq:B_ij}.
In what follows, this observation plays an important role.

\subsection{The spatial structure of quantities induced by the potential
forcing}

Let us assume that a quantity, say $A_{\BV K}$, is induced in the
inertial range by the forcing in \eqref{eq:KPhiW}. Here, by
``induced'', we do not necessarily mean that $A_{\BV K}$ equals the
forcing in \eqref{eq:KPhiW}, but that, first, it is present on the
same sets of $\BV K$ in the inertial range for which \eqref{eq:KPhiW}
is nonzero, and, second, its structure is largely governed by the
scaling properties of the Fourier transform $\phi_{\|\BV k\|}$ in the
inertial range.

First, let us consider the situation where $A_{\BV K}$ is nonzero
strictly in the subspaces $\cS_{ij}$ \eqref{eq:S_ij}, but otherwise
arbitrary. Here, we can assume that $\BV K\neq\BV 0$, since we are
interested in the inertial range. Let $a_{\BV K}$ be some function of
$\BV K$, and let us write
\begin{equation}
\label{eq:A_K}
A_{\BV K}=\sum_{i=1}^{N-1}\sum_{j=i+1}^N a_{\BV K} \delta(\BV k_i+\BV
k_j)\prod_{\myatop{m=1}{m\neq i,j}}^N\delta(\BV k_m).
\end{equation}
Clearly, $A_{\BV K}$ is zero almost everywhere, with the exception for
any $\BV K$ which belongs to one of the subspaces $\cS_{ij}$. In the
latter case, only one term in the sum above is nonzero (with indices
$i$ and $j$ corresponding to the given subspace $\cS_{ij}$), where
$A_{\BV K}$ is given via $a_{\BV K}$, scaled by the delta-functions.
Since $a_{\BV K}$ is arbitrary, we can say that the form of $A_{\BV
  K}$ in \eqref{eq:A_K} is the most general one which satisfies the
requirement of being nonzero in $\cS_{ij}$ and zero otherwise.  Next,
let us introduce the following notation for convenience:
\begin{equation}
a^{ij}_{\BV k}\eqdef\{a_{\BV K}:\BV K\in\cS_{ij}\},
\end{equation}
where $\BV k$ is the 3-dimensional wavevector parameter in the
definition of $\cS_{ij}$ in \eqref{eq:S_ij}. With this notation, we
can write \eqref{eq:A_K} as
\begin{equation}
\label{eq:A_K2}
A_{\BV K}=\sum_{i=1}^{N-1}\sum_{j=i+1}^N a^{ij}_{\BV k_i} \delta( \BV
k_i+\BV k_j)\prod_{\myatop{m=1}{m\neq i,j}}^N\delta(\BV k_m).
\end{equation}
Reverting the procedure in Appendix~\ref{app:potential_structure}, we
find that the inverse Fourier transform $A(\BV X)$ of $A_{\BV K}$,
whose form is specified in \eqref{eq:A_K}--\eqref{eq:A_K2}, is given
via
\begin{equation}
\label{eq:AX}
A(\BV X)=\sum_{i=1}^{N-1}\sum_{j=i+1}^N a^{ij}(\BV x_i-\BV x_j),
\end{equation}
where $a^{ij}(\BV x)$ is the 3-dimensional inverse Fourier transform
of $a^{ij}_{\BV k}$. We can see that, in the most general form, $A(\BV
X)$ depends on the differences of coordinates of all unordered pairs
of particles. One of the consequences of such a dependence is that
$A(\BV X)$ is invariant under a coordinate system shift in the
physical $3D$-space which is occupied by particles. Indeed, if we add
the same offset to the location $\BV x_i$ of each particle, then the
differences $\BV x_j-\BV x_i$ will remain the same, and thus $A(\BV
X)$ in \eqref{eq:AX} will not change.

Similarly, we can also presume that $A_{\BV K}$ is nonzero in the
bundles $\cB_{ij}$ of \eqref{eq:B_ij}. In this case, similarly to
\eqref{eq:KPhiW}, $A_{\BV K}$ can be expressed in the form of a
convolution of \eqref{eq:A_K}--\eqref{eq:A_K2} with some large scale
variable, whose spectral bandwidth is the same as that of $\BV W_{\BV
  K}$ in \eqref{eq:KPhiW}. In the physical space, this will become
\eqref{eq:AX}, multiplied by the inverse Fourier transform of that
large scale variable. If the large scale structures are filtered out,
the remainder will have the form in \eqref{eq:AX}.

Next, let us examine what happens to \eqref{eq:AX} in the framework of
conventional fluid dynamics. Recall that the Boltzmann equation
\cite{Abr17,Bol,CerIllPul} is obtained from the Liouville equation in
\eqref{eq:liouville} via the BBGKY formalism \cite{Bog,BorGre,Kir}, by
integrating the probability density $F$ over the coordinates and
velocities of all particles but one, obtaining the single-particle
marginal distribution $f_1$:
\begin{equation}
\label{eq:f1}
f_1(\BV x_1,\BV v_1)=\int_{\RR^{6(N-1)}}F(\BV X,\BV V)\,\dif\BV x_2
\ldots\dif\BV x_N\dif\BV v_2\ldots\dif\BV v_N.
\end{equation}
It is then presumed that all particles are distributed identically,
and thus the coordinates $\BV x_1$ and $\BV v_1$ of the first particle
are taken to be the coordinates of ``any'' particle, so that all
subscripts in \eqref{eq:f1} are dropped. The Boltzmann equation is the
corresponding transport equation for $f(\BV x,\BV v)$. The Euler and
Navier--Stokes equations are obtained by further integrating $f$
against various powers of $\BV v$, just as we have done above in
\eqref{eq:moments_liouville} for the complete system of particles.

If we apply the BBGKY procedure to \eqref{eq:AX} and integrate over
all coordinates but $\BV x_i$ for some $i$, $1\leq i\leq N$, it is
obvious that the resulting quantity will be a constant. Note, however,
that this happens solely due to the fact that the classical BBGKY
formalism is applied to a single particle. If the BBGKY formalism is
applied, say, to the joint distribution of two particles (that is, the
averaging is performed over all particles but two), then one
$3$-dimensional subspace $\cS_{ij}$ in \eqref{eq:S_ij} will be
preserved, which corresponds to this pair of particles. Similarly, if
the BBGKY formalism is applied to the joint distribution of $K$
particles, then $K(K-1)/2$ such subspaces will be preserved, each
corresponding to an unordered pair of these $K$ particles.

At this point, it is clear that the subspaces $\cS_{ij}$ in
\eqref{eq:S_ij}, and the corresponding bundles $\cB_{ij}$ in
\eqref{eq:B_ij}, are destroyed in the single-particle BBGKY formalism
of the conventional fluid dynamics, together with all quantities of
the form \eqref{eq:AX}, induced by the strong large scale flow in the
inertial range via the potential interaction. For comparison, let us
examine whether these quantities can be captured in the process of
measurement of an actual gas flow. Here, we assume that there is a
grid of $M$ measurement points, with coordinate offsets between them
given via $(\BV y_1,\ldots,\BV y_{M-1})$. We also assume that the
number of particles $N$ is much greater than the number of measurement
points $M$ (for example, $M\sim 10^3$, while $N\sim 10^{23}$). Each
measurement point contains a ``probe'' (that is, a measurement
device), which interacts with all particles which happen to pass
through that location, and measures their physical property of
interest.

In the course of the measurement, each probe registers the particles
which pass through that probe's location. Those particles which do not
pass through any of the probes, are not measured at all. Thus, the
measurement registers $M$ distinct sets of particles, the distances
between which given via $\BV y_m$. Since the probes do not discern
between individual particles, the quantity which will be recorded
constitutes the ensemble average over the particles, whose
coordinates, as well as the differences between them are known at the
time of the measurement:
\begin{equation}
\langle A\rangle(\BV y_m)=\text{ensemble average over all pairs }A(\BV
x_j-\BV x_i=\BV y_m).
\end{equation}
Thus, the offsets of measurement grid points $\BV y_m$ encode the
differences between particle's locations, which is what $A(\BV X)$ in
\eqref{eq:AX} depends on. Also, each pair of particles (say $i$th and
$j$th), registered at a location $m$, samples a single point from
their corresponding bundle $\cB_{ij}$ from \eqref{eq:B_ij}. However,
since the number of registered pairs (and, therefore, their respective
bundles $\cB_{ij}$) is large at any location $m$, the quantity which
is measured constitutes the ensemble average over these bundles, which
are effectively ``collapsed'' into the single $3$-dimensional space
sampled via $\BV y_m$.  While a lot of information is lost during this
collapse (for example, the phases of individual particle locations),
the bulk information, such as the power scaling of $A_{\BV K}$ shared
between all bundles $\cB_{ij}$, should remain largely preserved. It is
thus obvious that the subsequent discrete Fourier transformation (DFT)
of such a collapsed measurement will reveal the same power scaling as
shared across all bundles $\cB_{ij}$ in \eqref{eq:B_ij}. Thus, so far
we can summarize that:
\begin{enumerate}[\indent a.]
\item Via the interaction potential, the velocity of a strong large
  scale flow creates forcing in the inertial range, which is located
  in the bundles $\cB_{ij}$ given via \eqref{eq:B_ij}, which
  themselves consist of subspaces $\cS_{ij}$, given via
  \eqref{eq:S_ij};
\item In the conventional fluid dynamics, these bundles are destroyed
  by the BBGKY formalism \cite{Bog,BorGre,Kir}, which leads from the
  Liouville equation \eqref{eq:liouville} to the Boltzmann equation
  \cite{Abr17,Bol,CerIllPul};
\item The measurements of the flow, however, can register quantities
  which live in these bundles. While a measurement collapses these
  bundles into the measurement $3$-dimensional space, certain bulk
  properties can still manifest in the measured data.
\end{enumerate}
In what follows, we crudely estimate the scaling of the energy spectra
in the bundles $\cB_{ij}$, based on a realistic choice of the
interaction potential $\phi(r)$, and compare the estimates against
some known measurements.

\section{Weighted time averaging of the Liouville equation}
\label{sec:time_average}

Above, we found that a strong large scale flow induces a forcing at
the inertial scales via an interaction potential. This forcing,
however, lives in special bundles $\cB_{ij}$ of the form
\eqref{eq:B_ij}, which are shared by pairs of particles. Due to such a
structure, these bundles are not present in the equations of
conventional fluid dynamics, such as the Boltzmann, Euler and
Navier--Stokes equations, as they are destroyed in the process of the
BBGKY formalism \cite{Bog,BorGre,Kir}.  However, if an observation of
this flow is performed via a grid of measurement probes, then some
bulk properties of the flow shared across these bundles (such as, for
example, the power scaling with the Fourier wavenumber), can be
captured by the probes. This suggests that if a strong large scale
flow indeed causes the turbulent motions at the inertial scales by
means of the discovered forcing, then it would be interesting to
estimate, for example, the power scaling of the kinetic energy in
these bundles, and compare it with the observations.

Recall that turbulent kinetic energy spectra are observed for windowed
time averages of observed flows \cite{Obu62}. Thus, in order to find
out whether such spectra can manifest in the $N$-particle model
\eqref{eq:dyn_sys}, we need to transform the Liouville equation in
\eqref{eq:liouville} in such a way so that the resulting relations
describe either a windowed time average, or at least something similar
to it. Observe that \eqref{eq:liouville} is a linear PDE, and thus the
straightforward application of a windowed time average on both sides
transforms any instance of $F$ into the corresponding time average --
except for the time-derivative term, which becomes the difference
between the initial and terminal states of $F$. This is somewhat
inconvenient, since the terminal condition cannot simply be ignored or
set to zero (note that $F$ is a probability density, and thus it is
nonnegative and must integrate to 1).

Here, instead, to express the weighted time average of $F$, we will
use the quantity
\begin{equation}
\bar F(T,\BV X,\BV V)=\frac 1T\int_0^\infty F(t,\BV X,\BV V)e^{-t/T}
\dif t,
\end{equation}
where $T>0$ is a parameter. Observe that, for any $T$, $\bar F(T,\BV
X,\BV V)$ is also a probability density; indeed, first, $\bar F(T,\BV
X,\BV V)\geq 0$ (obviously), and, second,
\begin{equation}
\int\bar F(T,\BV X,\BV V)\dif\BV X\dif\BV V=\frac 1T\int_0^\infty
F(t,\BV X,\BV V)e^{-t/T} \dif t\dif\BV X\dif\BV V=\frac 1T
\int_0^\infty e^{-t/T} \dif t=1.
\end{equation}
While $\bar F(T,\BV X,\BV V)$ is not exactly the windowed time average
of $F$, it is qualitatively similar to the latter -- even though there
is no rigid cut-off at $T$ in the upper limit of the integral, a
``soft cut-off'' is present in the form of the exponential weight with
the characteristic scale $T$. Additionally, for large $T$, $\bar
F(T,\BV X,\BV V)$ indeed approaches the time average of $F$:
\begin{equation}
\label{eq:laplace_average}
\lim_{T\to\infty}\bar F(T,\BV X,\BV V)=\lim_{T\to\infty}\frac 1T
\int_0^T F(t,\BV X,\BV V)\dif t.
\end{equation}
The proof of \eqref{eq:laplace_average} can be found in \cite{GluMil};
we also provide it in Appendix \ref{app:proof} below.

Next, recall that the Laplace transform $\cL\{F\}(s,\BV X,\BV V)$ of
the probability density $F$ is given via
\begin{equation}
\cL\{F\}(s,\BV X,\BV V)=\int_0^\infty F(t,\BV X,\BV V)e^{-st}\dif t.
\end{equation}
Now, observe that, if we set $T=s^{-1}$, then $\bar F(T,\BV X,\BV V)$
is the same as $s\cL\{F\}(s,\BV X,\BV V)$:
\begin{equation}
\label{eq:F_hat}
\bar F(s^{-1},\BV X,\BV V)=s\int_0^\infty F(t,\BV X,\BV V)e^{-st} \dif
t=s\cL\{F\}(s,\BV X,\BV V).
\end{equation}
With the latter observation, it is easy to transform the Liouville
equation in \eqref{eq:liouville} into the corresponding equation for
$\bar F$. First, we apply the Laplace transformation on both sides of
the Liouville equation in \eqref{eq:liouville}:
\begin{equation}
s\cL\{F\}-F_0+\BV V\cdot\parderiv{}{\BV X}\cL\{F\}=\parderiv
\Phi{\BV X}\cdot\parderiv{}{\BV V}\cL\{F\},
\end{equation}
where $F_0=F(0,\BV X,\BV V)$ is the initial condition for $F$.  Then,
multiplying both sides by $s$ and denoting $T=s^{-1}$, we arrive at
\begin{equation}
\label{eq:liouville_laplace}
\frac 1T(\bar F-F_0)+\BV V\cdot\parderiv{\bar F}{\BV X}=\parderiv\Phi{
  \BV X}\cdot\parderiv{\bar F}{\BV V}.
\end{equation}
Since $\bar F$ is a probability density, the same techniques can be
applied for \eqref{eq:liouville_laplace}, as above for
\eqref{eq:liouville}.  Denoting the corresponding Laplace-averaged
velocity moments for $\bar F$ via $\brho$, $\bU$, $\bE$, $\bQ$ and
$\bR$, we obtain the following Laplace-averaged moment equations:
\begin{subequations}
\label{eq:moments_laplace}
\begin{equation}
\frac 1T(\brho-\rho_0)+\parderiv{}{\BV X}\cdot(\brho\bU)=0, \qquad
\frac 1T(\brho\bU-\rho_0\BV U_0)+\parderiv{}{\BV X}\cdot(\brho\bE)=
-\brho\parderiv\Phi{\BV X},
\end{equation}
\begin{equation}
\frac 1T(\brho\bE-\rho_0\BM E_0)+\parderiv{}{\BV
  X}\cdot(\brho(\bQ+\bR))=-\brho\bigg(\parderiv\Phi{\BV
  X}\bU^T+\bU\parderiv \Phi{\BV X}^T\bigg).
\end{equation}
\end{subequations}
Further applying the Fourier transformation in $\BV X$ to
\eqref{eq:moments_laplace} in the same manner as was done in
\eqref{eq:moments_fourier}, we arrive at
\begin{subequations}
\label{eq:moments_laplace_fourier}
\begin{equation}
\frac 1{\hi T}(\brho_{\BV K}-\rho_{0,\BV K})+\BV K\cdot(\brho_{\BV K}
\ast\bU_{\BV K})=0,
\end{equation}
\begin{equation}
\frac 1{\hi T}(\brho_{\BV K}\ast\bU_{\BV K}-\rho_{0,\BV K}\ast\BV
U_{0,\BV K})+\BV K\cdot(\brho_{\BV K}\ast\bE_{\BV K})=-\brho_{\BV K}
\ast(\BV K\Phi_{\BV K}),
\end{equation}
\begin{multline}
\frac 1{\hi T}(\brho_{\BV K}\ast\bE_{\BV K}-\rho_{0,\BV K}\ast\BM
E_{0,\BV K})+\BV K\cdot(\brho_{\BV K}\ast(\bQ_{\BV K}+\bR_{\BV K}))=
\\=-\brho_{\BV K}\ast\left[(\BV K\Phi_{\BV K})\ast\bU_{\BV K}^T
  +\bU_{\BV K} \ast (\BV K\Phi_{\BV K})^T\right].
\end{multline}
\end{subequations}
For the same assumptions as in \eqref{eq:moments_fourier_simplified},
in the inertial range we obtain
\begin{subequations}
\label{eq:moments_laplace_fourier_simplified_2}
\begin{equation}
\frac 1{\hi T}\brho_{\BV K}+\BV K\cdot(\brho_{\BV K}\ast\bU_{\BV K})
=0,
\qquad
\frac 1{\hi T}\brho_{\BV K}\ast\bU_{\BV K}+\BV K\cdot(\brho_{\BV K}
\ast\bE_{\BV K})=\BV 0,
\end{equation}
\begin{equation}
\frac 1{\hi T}\brho_{\BV K}\ast\bE_{\BV K}+\BV K\cdot(\brho_{\BV K}
\ast(\bQ_{\BV K}+\bR_{\BV K}))=-\brho_{\BV K}\ast\left[(\BV K\Phi_{\BV
    K})\ast\bW_{\BV K}^T+\bW_{\BV K}\ast(\BV K\Phi_{\BV K})^T\right],
\end{equation}
\end{subequations}
where $\bW_{\BV K}$ is the Laplace average of the strong large scale
velocity $\BV W_{\BV K}$ in \eqref{eq:moments_fourier_simplified}, and
we recall that the initial values $\rho_{0,\BV K}$, $\BV U_{0,\BV K}$
and $\BM E_{0,\BV K}$ in the inertial range are presumed to be zero.

\section{Simplified relations and scaling estimates for the inertial
  range}
\label{sec:simplified}

In what follows, our goal is to estimate bulk power scaling relations
in the bundles $\cB_{ij}$ from \eqref{eq:B_ij} between the Fourier
transform $\phi_{\|\BV k\|}$ of the interaction potential and the
Fourier transform $\bE_{\BV K}$ of the Laplace-averaged kinetic
energy, and compare the estimates with the observations. Here, we must
keep in mind that what is observed is not necessarily the exact
solution, but rather the most ``visible'' component of the solution.
For example, when a power scaling of the kinetic energy is observed,
it does not necessarily mean that the solution has a precise power
scaling form -- but rather that its other components may decay faster
than that, and thus be less ``visible''. Additionally, it is known
that, in a large variety of configurations of the large scale flow,
the turbulent motions do not manifest themselves at all (this is known
as the ``laminar'' flow). Therefore, for the purpose of this work,
further we will tacitly assume that the large scale flow configuration
is such that the turbulent motions arise, and will focus specifically
on the estimates which yield the bulk power scaling of the kinetic
energy spectra, while keeping in mind that other, more rapidly
decaying, components of the solution could also be ``invisibly''
present.

Since we do not need to look for exact solutions of
\eqref{eq:moments_laplace_fourier_simplified_2}, we can instead
identify dominant terms in
\eqref{eq:moments_laplace_fourier_simplified_2} and crudely estimate
the order-of-magnitude relations between them. For that, we make the
following simplifying assumptions in
\eqref{eq:moments_laplace_fourier_simplified_2}:
\begin{enumerate}[\indent a.]
\item On the chosen averaging time scale $T$, the flow is effectively
  incompressible, that is, $\brho_{\BV K}$ is confined to a very
  narrow wavenumber range $\|\BV K\|\ll 1$, and thus can be factored
  out of the convolution integrals -- in fact, this is what is
  observed in practice for many turbulent flows;
\item On the chosen averaging time scale $T$, the skewness moment
  $\bQ_{\BV K}$ in the inertial range can be neglected in comparison
  with the combination of the velocity and energy moments $\bR_{\BV
    K}$ -- as we mentioned above, in practical scenarios, the skewness
  and higher-order non-Gaussian moments typically manifest themselves
  at viscous scales, while in the inertial range the non-Gaussian
  moments are small;
\item In the energy equation, the Fourier transform $\bE_{\BV K}$
  decays faster with $\BV K$ than the remaining part of the advection
  term $\BV K\cdot\bR_{\BV K}$, and thus can be neglected in the
  inertial range. This assumption will be shown to hold for all
  estimates below.
\end{enumerate}
The resulting simplified relations for the velocity and energy are
given via
\begin{equation}
\label{eq:moments_laplace_fourier_simplified}
\bU_{\BV K}\approx -\hi T\BV K\cdot\bE_{\BV K},\qquad\BV K\cdot
\bR_{\BV K}\approx-(\BV K\Phi_{\BV K})\ast \bW_{\BV K}^T-\bW_{\BV K}
\ast(\BV K\Phi_{\BV K})^T.
\end{equation}
Below, we will assume that the scaling of the Fourier transforms
$\bU_{\BV K}$ and $\bE_{\BV K}$ in the bundles $\cB_{ij}$ from
\eqref{eq:B_ij} with $\BV k$ (where $\BV k$ is the wavevector
parameter in \eqref{eq:S_ij}) is governed by these approximate
relations in \eqref{eq:moments_laplace_fourier_simplified}.

Next, since the nonzero values of the Laplace-averaged strong large
scale flow velocity $\bW_{\BV K}$ are confined to a small ball $\|\BV
K\|<b_{\BV W}$, we observe that, in any bundle $\cB_{ij}$ from
\eqref{eq:B_ij} the scaling of $\bR_{\BV K}$ should be the same as the
scaling of $\phi_{\|\BV k\|}$, with $\BV K$ within $\cB_{ij}$ in the
inertial range being roughly of the form in \eqref{eq:S_ij}:
\begin{equation}
\label{eq:R_scaling}
\BV K\cdot\bR_{\BV K}\sim\BV K\phi_{\|\BV k\|},\quad\text{or}\quad
\bR_{\BV K}\sim\phi_{\|\BV k\|}.
\end{equation}
Recall that $\bR_{\BV K}$ itself consists of convolutions of $\bU_{\BV
  K}$ with itself and $\bE_{\BV K}$. Thus, even though the scaling of
$\bR_{\BV K}$ in a bundle $\cB_{ij}$ is fully determined via
\eqref{eq:R_scaling}, the convolutions $\bU_{\BV K}\circledast\bE_{\BV
  K}$ and $\bU_{\BV K}\circledast\bU_{\BV K}\circledast\bU_{\BV K}$
can, in principle, combine the contributions from different bundles.
Besides, even though $\bR_{\BV K}$ is zero when $\BV K\notin\cB_{ij}$,
it does not mean that $\bU_{\BV K}$ and $\bE_{\BV K}$ are necessarily
zero for the same $\BV K$.  Ultimately, not every flow is necessarily
turbulent, even if it does have a strong large scale velocity
component.

Therefore, in what follows, we will consider the solutions $\bU_{\BV
  K}$ and $\bE_{\BV K}$ of
\eqref{eq:moments_laplace_fourier_simplified_2} and
\eqref{eq:moments_laplace_fourier_simplified}, which have the same
bulk structure as that of $\bR_{\BV K}$ (that is, they are nonzero in
the bundles $\cB_{ij}$ and zero otherwise). Additionally, we will
assume that the values of $\bR_{\BV K}$ in a bundle $\cB_{ij}$ are
largely determined by the values of $\bU_{\BV K}$ and $\bE_{\BV K}$
{\em in that same bundle}, that is, the contribution of convolutions
across distinct bundles is either negligible, or at least does not
affect the scaling significantly. To make the scaling estimates, we
will also assume that the double velocity convolution in the formula
for $\bR_{\BV K}$ in \eqref{eq:RK} scales in the same manner as
$\bR_{\BV K}$ itself in \eqref{eq:R_scaling},
\begin{equation}
\label{eq:UR_scaling}
\bU_{\BV K}\circledast\bU_{\BV K}\circledast\bU_{\BV K}\sim\bR_{\BV
  K}\sim\phi_{\|\BV k\|},
\end{equation}
in any bundle $\cB_{ij}$. Just as the assumption about the energy
decay above, this assumption will be shown to hold for all estimates
below.

Given the sparse structure of $\Phi_{\BV K}$ in \eqref{eq:PhiK},
observe that the convolution in \eqref{eq:KPhiW} qualitatively
replicates the pattern of the strong large scale flow velocity
$\bW_{\BV K}$ along the bundles $\cB_{ij}$ in \eqref{eq:B_ij}.  While
this pattern is not copied verbatim (as the term $\BV K\Phi_{\BV K}$
adjusts the scaling with $\BV K$), it is clear that some qualitative
degree of ``self-similarity'' must be present in \eqref{eq:KPhiW}
across the inertial range. If the relation in \eqref{eq:UR_scaling}
holds, then one can expect such self-similar patterns to also manifest
in the inverse Fourier transform of the Laplace-averaged velocity
$\bU_{\BV K}$. This is what seems to be usually observed in turbulent
flows.

In order to make the scaling estimates of the Fourier transforms of
the velocity and energy, below we separate the inertial range into the
following two subranges:
\begin{enumerate}[\indent a.]
\item {\bf Large scale inertial subrange} -- this is the subrange of
  the Fourier wavenumbers adjacent to the large scale flow;
\item {\bf Small scale inertial subrange} -- this is the subrange of
  the Fourier wavenumbers adjacent to the viscous range.
\end{enumerate}
This separation must be put in place due to the presence of the
velocity cut-off in the viscous spatial range -- as we show below, the
scaling estimates must be computed differently, depending on how far
away the wavenumber is from the viscous velocity cut-off.

\subsection{Scaling estimate for the large scale inertial subrange}

Above, we assumed that a physically relevant solution of
\eqref{eq:moments_fourier} has a cut-off for the Fourier transform of
the velocity at the boundary of the viscous range. However, for a
Fourier wavenumber $\BV K$ which is in the inertial range, but in the
proximity of the large scale flow, the effect of the viscous cut-off
can be very small, since it happens on a much smaller scale. In such a
case, one can assume that the viscous cut-off is not present, and,
therefore, to estimate the scaling of the velocity and energy in the
inertial range near the large scale flow, we can use the standard
regularity estimates from the Fourier analysis.

Let us write the expression for the double convolution explicitly via
the corresponding integral:
\begin{equation}
\label{eq:UUU}
\bU_{\BV K}\circledast\bU_{\BV K}\circledast\bU_{\BV K}=\int_{
  \RR^{3N}}\int_{\RR^{3N}}\bU_{\BV K-\BV K'}\otimes\bU_{\BV K'-\BV
  K''} \otimes\bU_{\BV K''} \dif\BV K''\dif\BV K'.
\end{equation}
Clearly, if $\|\BV K\|$ is much smaller than the threshold for the
viscous cut-off, then, even in the absence of the cut-off, due to
power decay, the contribution from the small scale wavenumbers is much
smaller than from those which surround $\BV K$. Subsequently, the
result of the double convolution in \eqref{eq:UUU} without the viscous
cut-off should not be much different from that with the cut-off
present.

In such a situation, we can ignore the velocity cut-off at viscous
scales, and proceed as if the scaling of $\bU_{\BV K}$ and $\bR_{\BV
  K}$ above in \eqref{eq:R_scaling} and \eqref{eq:UR_scaling} extends
onto all Fourier wavenumbers. This scaling, in turn, determines the
regularity class for $\bR$ itself, which is the inverse Fourier
transform of $\bR_{\BV K}$. At the same time, the convolution in the
Fourier space becomes the product in the physical space:
\begin{equation}
\bR\sim\bU\otimes\bU\otimes\bU.
\end{equation}
Since $\bR$ is, effectively, a cubic power of $\bU$, it means that
$\bU$ belongs to the same regularity class as $\bR$. This, in turn,
means that, for $\|\BV K\|$ in the inertial range near the large
scales, $\bU_{\BV K}$ should scale in the same manner as $\bR_{\BV K}$
along any bundle $\cB_{ij}$:
\begin{equation}
\label{eq:u_large_scale}
\bU_{\BV K}\sim\bR_{\BV K}\sim\phi_{\|\BV k\|}.
\end{equation}
However, due to \eqref{eq:moments_laplace_fourier_simplified}, the
latter means that the scaling for the energy $\bE_{\BV K}$ is given
via
\begin{equation}
\label{eq:e_large_scale}
\bE_{\BV K}\sim\frac{\phi_{\|\BV k\|}}{\|\BV k\|},
\end{equation}
along any bundle $\cB_{ij}$ from \eqref{eq:B_ij}. It remains to be
verified that the obtained scaling does not violate the assumptions in
\eqref{eq:moments_laplace_fourier_simplified} and
\eqref{eq:UR_scaling}.  Indeed, observe that, since $\bE$ is in the
``better'' regularity class than $\bU$, the product $\bU\otimes\bE$
(which is what becomes of convolutions $\bU_{\BV K}\circledast\bE_{\BV
  K}$ in the coordinate space) falls into the same regularity class as
$\bU$ itself. Thus, we conclude that all terms in $\bR_{\BV K}$ scale
in the same fashion within the inertial range near the large scale
flow, which ascertains the validity of \eqref{eq:UR_scaling}.  Also,
$\bE_{\BV K}$ indeed decays faster with $\BV K$ than $\BV
K\cdot\bR_{\BV K}$, so that
\eqref{eq:moments_laplace_fourier_simplified} remains valid.

\subsection{Scaling estimate for the small scale inertial subrange}

If the wavenumber $\BV K$ in
\eqref{eq:moments_laplace_fourier_simplified} is in the vicinity of
the viscous velocity cut-off, the regularity estimates above cannot be
used. The reason for this is that the velocity convolutions in
$\bR_{\BV K}$ have a limited effective domain of integration, which
completely changes the scaling behavior of the convolvant. To
illustrate this, let us shift the dummy variables of integration in
\eqref{eq:UUU} as follows:
\begin{equation}
  \BV K'\to\BV K'+2\BV K/3,\qquad\BV K''\to\BV K''+\BV K/3.
\end{equation}
This change of the variables of integration leads to
\begin{equation}
\label{eq:convolution_shifted}
\bU_{\BV K}\circledast\bU_{\BV K}\circledast\bU_{\BV K}=\int_{\RR^{3N}
}\int_{\RR^{3N}}\bU_{\BV K/3-\BV K'}\otimes\bU_{\BV K/3 +\BV K'-\BV
  K''}\otimes\bU_{\BV K/3+\BV K''}\dif\BV K''\dif\BV K'.
\end{equation}
Now it is easy to see that if $\BV K/3$ above is chosen near the
viscous velocity cut-off, then the values of $\BV K'$ and $\BV K''$,
for which the integrand is nonzero, are clustered in a small region
around zero. In order to see how the power scaling of the convolution
is affected, let us assume that $\bU_{\BV K}$ has the form
\begin{equation}
\label{eq:UY}
\bU_{\BV K}=\|\BV K\|^{-\gamma}\BV Y_{\BV K},
\end{equation}
where $\gamma>0$ is a power scaling constant, while $\BV Y_{\BV K}$
scales as $O(1)$ in the inertial range, and rapidly decays to zero
beyond the viscous cut-off. Upon substitution of \eqref{eq:UY} into
\eqref{eq:convolution_shifted}, we arrive at
\begin{multline}
\bU_{\BV K}\circledast\bU_{\BV K}\circledast\bU_{\BV K}=\int_{\RR^{3N}
}\int_{\RR^{3N}}\left(\|\BV K/3-\BV K'\|\|\BV K/3+\BV K'-\BV K''\|
\|\BV K/3+\BV K''\|\right)^{-\gamma}\\\BV Y_{\BV K/3-\BV K'}\otimes\BV
Y_{\BV K/3+\BV K'-\BV K''}\otimes\BV Y_{\BV K/3 +\BV K''}\dif\BV K''
\dif\BV K'.
\end{multline}
At this point, it is already clear that, if $\BV K'$ and $\BV K''$
above are confined to a small region around zero, the product of norms
above can be expected to scale as $\|\BV K\|^{-3\gamma}$. To elaborate
more on this estimate, we first write, for the product of norms alone,
\begin{multline}
\|\BV K/3-\BV K'\|\|\BV K/3+\BV K'-\BV K''\|\|\BV K/3+\BV K''\|
=\\=\frac{\|\BV K\|^3}{27}\Big[\Big( 1-\frac{6\BV K^T\BV K'}{\|\BV
    K\|^2} +\frac{9\|\BV K'\|^2}{\|\BV K\|^2}\Big)\Big( 1+\frac{6\BV
    K^T(\BV K'-\BV K'')}{\|\BV K\|^2}+\frac{9\|\BV K'-\BV
    K''\|^2}{\|\BV K\|^2}\Big)\\\Big(1+\frac{6\BV K^T\BV K''}{\|\BV
    K\|^2}+ \frac{9\|\BV K''\|^2}{\|\BV
    K\|^2}\Big)\Big]^{1/2}=\frac{\|\BV
  K\|^3}{27}\sqrt{G\left(\frac{2\BV K}{\|\BV K\|},\frac{3\BV K'}{\|\BV
    K\|},\frac{3\BV K''}{\|\BV K\|}\right)},
\end{multline}
where $G(\BV a,\BV b,\BV c)$ is given via
\begin{multline}
  G(\BV a,\BV b,\BV c)=(1-\BV a^T\BV b+\|\BV b\|^2)(1+\BV a^T(\BV
  b-\BV c)+\|\BV b-\BV c\|^2) (1+\BV a^T\BV c+\|\BV c\|^2)=\\= 1+\|\BV
  b\|^2+\|\BV c\|^2+\|\BV b-\BV c\|^2+(\BV b-\BV a)^T\BV b(\BV a+\BV
  c)^T\BV c+(\BV b-\BV a)^T\BV b(\BV a+\BV b-\BV c)^T(\BV b-\BV c)
  +\\+(\BV a+\BV b-\BV c)^T(\BV b-\BV c)(\BV a+\BV c)^T\BV c+(\BV b
  -\BV a)^T\BV b(\BV a+\BV b-\BV c)^T(\BV b-\BV c)(\BV a+\BV c)^T\BV c,
\end{multline}
and notably lacks the linear terms in $\BV b$ and $\BV c$, since they
cancel out due to the alternating signs. The form of $G$ above results
in the order-of-magnitude behavior
\begin{equation}
G\left(\frac{2\BV K}{\|\BV K\|},\frac{3\BV K'}{\|\BV K\|},\frac{3\BV
  K''}{\|\BV K\|}\right)=1+O\left(\frac{\|\BV K'\|^2}{\|\BV K\|^2}
\right)+O\left(\frac{\|\BV K''\|^2}{\|\BV K\|^2} \right)
+O\left(\frac{\|\BV K\|\|\BV K''\|}{\|\BV K\|^2}\right),
\end{equation}
that is, for $\|\BV K'\|\sim\|\BV K''\|\ll\|\BV K\|$, $G$ behaves
similarly to a constant, with at least a quadratic correction term.
The convolution thus becomes
\begin{multline}
\label{eq:Y_integral}
\bU_{\BV K}\circledast\bU_{\BV K}\circledast\bU_{\BV K}=\frac{
  27^\gamma}{\|\BV K\|^{3\gamma}}\int_{\RR^{3N}}\int_{\RR^{3N}}
G\left(\frac{2\BV K}{\|\BV K\|},\frac{3\BV K'}{\|\BV K\|},\frac{3\BV
  K''}{\|\BV K\|}\right)^{-\gamma/2}\\\BV Y_{\BV K/3-\BV K'}\otimes\BV
Y_{\BV K/3+\BV K'-\BV K''}\otimes\BV Y_{\BV K/3 +\BV K''}\dif\BV K''
\dif\BV K'.
\end{multline}
Even though $G$ behaves like a constant, and $\BV Y_{\BV K}$ is
$O(1)$, the full integral above does not necessarily scale like a
constant in $\BV K$. This happens because the size of the effective
domain of integration depends on $\BV K$ too -- indeed, the farther
away is $\|\BV K\|$ from the viscous cut-off, the larger is the
effective domain of integration (whose size scales as a power of the
difference between $\|\BV K\|$ and the viscous cut-off). If the
integral scales proportionally to the effective domain of the
integration, on a log-log plot, such a dependence will look like a
rapidly, superlinearly decreasing function of $\|\BV K\|$, and thus
the power scaling in $\|\BV K\|$ cannot be achieved in such a case.

Therefore, in order to attain the power scaling in $\|\BV K\|$, the
magnitude of the integral in \eqref{eq:Y_integral} should not scale
with the size of the domain of integration. This can be achieved if
$\BV Y_{\BV K}$ consists of self-similar patterns with alternating
signs, so that appropriate cancellations can occur during the
integration.  Such an ansatz seems to be supported by observations,
where turbulent structures usually consist of chaotically oriented
self-similar eddies spanning multiple scales. Also, since the
structure of $\BV Y_{\BV K}$ is determined by the structure of the
strong large scale flow $\bW_{\BV K}$, the latter should likely have a
suitable pattern to induce a turbulent flow -- clearly, not all large
scale flows necessarily produce turbulent effects. Therefore, a
naturally emerging (and likely difficult) problem here is to determine
a variety of patterns for $\bW_{\BV K}$ which can produce appropriate
forcings via \eqref{eq:KPhiW}, so that the resulting velocity fields
have the suitable self-canceling structure.

At this point, what remains to be determined is the corresponding
energy scaling. Clearly, if $\BV K$ belongs to a bundle $\cB_{ij}$ in
\eqref{eq:B_ij}, then so does $\BV K/3$. Thus, combining
\eqref{eq:UR_scaling}, \eqref{eq:UY} and \eqref{eq:Y_integral}, we
arrive at
\begin{equation}
\label{eq:u_small_scale}
\|\BV K\|^{-3\gamma}\sim\phi_{\|\BV k\|},\quad\text{and, therefore,}
\quad\bU_{\BV K}\sim\|\BV K\|^{-\gamma}\sim\sqrt[3]{\phi_{\|\BV k\|}},
\end{equation}
where $\BV k$ is the wavevector which parameterizes the bundle
$\cB_{ij}$ in \eqref{eq:B_ij} via \eqref{eq:S_ij}. Due to
\eqref{eq:moments_laplace_fourier_simplified}, the latter means that,
in the small scale inertial range near the viscous velocity cut-off,
the energy scales as
\begin{equation}
\label{eq:e_small_scale}
\bE_{\BV K}\sim\frac{\sqrt[3]{\phi_{\|\BV k\|}}}{\|\BV k\|}.
\end{equation}
Lastly, we have to verify that, for the obtained scaling, the
assumptions above in \eqref{eq:moments_laplace_fourier_simplified} and
\eqref{eq:UR_scaling} indeed hold. Here, since $\BV K/3$ is chosen to
be near the viscous cut-off for $\bU_{\BV K}$, and $\bE_{\BV K}$ is
given via \eqref{eq:moments_laplace_fourier_simplified} (and is a
``multiple'' of the velocity), then the convolution of $\bU_{\BV K}$
with $\bE_{\BV K}$ at most reaches the wavenumber $2\BV K/3$, and
never $\BV K$ itself. Thus, around $\BV K$, $\bR_{\BV K}$ consists
only of the double velocity convolution, which ascertains
\eqref{eq:UR_scaling}.  Similarly, the energy $\bE_{\BV K}$ alone can
never extend beyond $\BV K/3$, which means that the only quantity
affected by the forcing near $\BV K$ in
\eqref{eq:moments_laplace_fourier_simplified} is $\bR_{\BV K}$.

\section{The choice of the interaction potential and the scaling
estimates}

Above in \eqref{eq:u_large_scale}, \eqref{eq:e_large_scale},
\eqref{eq:u_small_scale} and \eqref{eq:e_small_scale}, we obtained the
estimates for the velocity and energy scaling in the inertial range,
which depend on the scaling of the Fourier transform $\phi_{\|\BV
  k\|}$ of the interaction potential $\phi(r)$. Here, we need to
choose the form of $\phi(r)$, which would provide a realistic Fourier
transform in \eqref{eq:phik}. The main problem here is that, in
reality, the interactions between molecules are largely governed by
quantum-mechanical effects (in particular, the Pauli exclusion
principle). An ``interaction potential'' is merely the averaged,
statistical manifestation of the latter in the classical limit, and,
therefore, no potential fully describes the interactions between
real-world molecules.

Typically, the choice of a potential is defined by the context of a
relevant problem. For example, the widely known Buckingham \cite{Buc}
and Lennard-Jones \cite{Len} potentials are constructed using
semi-empirical reasoning for the dynamics of low-energy, densely
packed atoms primarily in liquids, and do not generally provide an
accurate description outside of this context. In particular, the main
focus of both the Lennard-Jones and Buckingham potentials is on the
accuracy of the attracting term, which is known, from experiments and
observations, to scale as $\sim r^{-6}$ with the distance. On the
other hand, the repulsion terms of the aforementioned potentials,
which are important for the high-energy collisions in a ``normal''
gas, are chosen largely out of convenience. As a result, not only the
Fourier integrals of both the Lennard-Jones and Buckingham potentials
are unbounded due to near-zero singularities, but even if they were
somehow made bounded (say, via a suitable regularization limit), they
would have been effectively unrelated.

In the present context, our goal is to choose the potential model
which offers a realistic scaling of the Fourier transform.  This means
that the qualitative, bulk behavior of the potential must be
physically reasonable, instead of being quantitatively tailored to a
particular narrow range of scales and energies. At the same time, the
model must be sufficiently simple, to allow explicit treatment without
the need to resort to numerical simulations. Arguably, the simplest
model of this kind is the Thomas--Fermi model \cite{Tho,Fer}, which is
based upon a (rather crude) quantum-mechanical formulation of the
atomic structure via Schr\"odinger's equation. According to the
Thomas--Fermi model, the potential $\phi(r)$ has the form
\begin{equation}
\label{eq:thomas-fermi}
\phi(r)=\frac{\phi_0}r\eta\left(\frac r\sigma\right),
\end{equation}
where $\phi_0$ is a dimensional constant, $\eta(r)$ is a screening
function, and $\sigma$ is the characteristic screening distance. The
screening function $\eta(r)$ satisfies the Thomas--Fermi nonlinear
differential equation:
\begin{equation}
\deriv{^2\eta}{r^2}=\sqrt{\frac{\eta^3}r},\qquad\eta(0)=1,
\quad\eta(\infty)=0.
\end{equation}
Note that the solution to the Thomas--Fermi equation is unavailable in
the explicit form. However, in order to estimate the scaling of the
Fourier transform $\phi_{\|\BV k\|}$, we only need to know certain
properties of the solution. It is known \cite{KobMatNagUme,Som} that,
for $0\leq r<\infty$, $\eta(r)$ is a bounded strictly positive
monotonically decreasing function, which behaves asymptotically as
$\sim 144r^{-3}$ for $r\to\infty$. It also possesses continuous
bounded strictly negative monotonically increasing derivative, whose
initial value $\eta'(0)\approx -1.588$. Thus, the Thomas--Fermi
potential itself behaves as $\sim r^{-1}$ for $r\ll 1$, and $\sim
r^{-4}$ for $r\gg 1$.

\subsection{The scaling for the Thomas--Fermi potential}

Substituting \eqref{eq:thomas-fermi} into \eqref{eq:phik}, and
rescaling $r\to\sigma r$, we obtain
\begin{equation}
\label{eq:etak}
\phi_{\|\BV k\|}=\frac{2\phi_0\sigma}{\|\BV k\|}\int_0^\infty\eta(r)
\sin(2\pi\sigma\|\BV k\|r)\dif r.
\end{equation}
Above, the integral alone is the Fourier sine transform of a
continuously differentiable function on the right half-axis with a
discontinuous odd extension. Therefore, it scales as $(\sigma\|\BV
k\|)^{-1}$ as $\|\BV k\|\to\infty$, which, in turn, means that the
Fourier transform $\phi_{\|\BV k\|}$ in \eqref{eq:etak} should scale
as $\|\BV k\|^{-2}$ as $\|\BV k\|\to\infty$.

For a more accurate estimate, let us first note that, for $\sigma\|\BV
k\|\ll 1$ (which corresponds to large scales) we can assume that
$\eta(r)$ decays to zero much faster than the scale of variation of
the sine function. In such a case, we can truncate the sine function
to its own leading order Taylor term, which yields the estimate
\begin{equation}
\phi_{\|\BV k\|}=\frac{2\phi_0\sigma}{\|\BV k\|}\int_0^\infty\eta(r)
\sin(2\pi\sigma\|\BV k\|r)\dif r\approx 4\pi\phi_0\sigma^2
\int_0^\infty r\eta(r)\dif r\sim\sigma^2,
\end{equation}
that is, at large scales the Fourier transform of the Thomas--Fermi
potential is a small constant. This, of course, is to be expected,
since it is known that the effect of interatomic potentials is very
limited at large scales. For the velocity and energy scaling in
\eqref{eq:u_large_scale} and \eqref{eq:e_large_scale} in any bundle
$\cB_{ij}$ at large inertial scales, this formally means that
\begin{equation}
\bU_{\BV K}\sim 1,\qquad\bE_{\BV K}\sim\|\BV k\|^{-1}.
\end{equation}
However, in practice the Thomas--Fermi potential would be unable to
affect the dynamics at large scales, due to its short range (or, to
put it more formally, it would not be practically possible to choose
the strength of the large scale flow $\bW_{\BV K}$ so that the
approximate relations in \eqref{eq:moments_laplace_fourier_simplified}
would hold).

For $\sigma\|\BV k\|\sim 1$ (that is, at small scales), the estimate
above becomes invalid. Instead, one can integrate \eqref{eq:etak} by
parts to obtain
\begin{equation}
\label{eq:etak2}
\phi_{\|\BV k\|}=\frac{\phi_0}{\pi\|\BV k\|^2}\bigg(1+\int_0^\infty
\eta'(r)\cos(2\pi\sigma\|\BV k\|r)\dif r\bigg)\sim\|\BV k\|^{-2},
\end{equation}
which means that at small scales the Fourier transform of the
Thomas--Fermi potential behaves as the inverse square of the
wavenumber. For the velocity and energy scaling in
\eqref{eq:u_small_scale} and \eqref{eq:e_small_scale} in any bundle
$\cB_{ij}$ at small inertial scales, this formally means that
\begin{equation}
\label{eq:tf_2}
\bU_{\BV K}\sim\|\BV k\|^{-2/3},\qquad\bE_{\BV K}\sim\|\BV k\|^{-5/3}.
\end{equation}
Observe that the estimated energy scaling coincides with the famous
Kolmogorov scaling
\cite{Kol41a,Kol41b,Kol41c,Obu41,Obu49,Cha49,Cor51}.  Also, note that
any potential of the form \eqref{eq:thomas-fermi}, whose screening
function has a continuous bounded derivative -- for example, the
Ziegler--Biersack--Littmark potential \cite{ZieBieLit} -- will produce
the same scaling as in \eqref{eq:tf_2}.

\subsection{The scaling for the electrostatic potential}

From the relation in \eqref{eq:etak2}, it is easy to obtain the
explicit formula for the Fourier transform of the electrostatic
potential. Namely, observe that the electrostatic $1/r$-potential can
be obtained from the Thomas--Fermi potential in \eqref{eq:tf} by
taking the limit $\sigma\to\infty$. Applying the same limit in
\eqref{eq:etak2} for $\|\BV k\|>0$ yields
\begin{equation}
\lim_{\sigma\to\infty}\int_0^\infty\eta'(r)\cos(2\pi\sigma\|\BV k\|r)
\dif r=\lim_{\omega\to\infty}\int_0^\infty\eta'(r)\cos(\omega r)\dif
r=0,
\end{equation}
since $\eta'(r)$ is continuous and bounded, and thus its cosine
Fourier transform decays to zero with increasing wavenumber. This
leads to
\begin{equation}
\phi_{\|\BV k\|}^{el}=\frac{\phi_0^{el}}{\pi\|\BV k\|^2}.
\end{equation}
As we can see, the Fourier transform for the electrostatic potential
decays as $\|\BV k\|^{-2}$ throughout the whole range of Fourier
wavenumbers, including the large scales. For the velocity and energy
scalings in \eqref{eq:u_large_scale}, \eqref{eq:e_large_scale},
\eqref{eq:u_small_scale} and \eqref{eq:e_small_scale} in any bundle
$\cB_{ij}$, this means
\begin{subequations}
\label{eq:el_2}
\begin{equation}
\bU_{\BV K}\sim\|\BV k\|^{-2},\qquad\bE_{\BV K}\sim\|\BV k\|^{-3}
\qquad\text{at large inertial scales},
\end{equation}
\begin{equation}
\bU_{\BV K}\sim\|\BV k\|^{-2/3},\qquad\bE_{\BV K}\sim\|\BV k\|^{-5/3}
\qquad\text{at small inertial scales}.
\end{equation}
\end{subequations}
Observe that, at small inertial scales, the energy scaling for the
electrostatic potential is the same as that for the Thomas--Fermi
potential in \eqref{eq:tf_2}, and also coincides with the
famous Kolmogorov scaling
\cite{Kol41a,Kol41b,Kol41c,Obu41,Obu49,Cha49,Cor51}.

\section{Comparison with observations}

Above, we estimated the scaling of the $3N\times 3N$-dimensional
matrix $\bE_{\BV K}$, which is the Fourier transform of the full
Laplace-averaged kinetic energy matrix $\bE(\BV X)$, respectively.
The energy matrix contains the full set of velocity moments of the
second order for all particles, as functions of all their distinct
locations.

Clearly, such a quantity cannot be measured in a realistic experiment
or an observation. As described above in Section \ref{sec:structures},
a typical experiment or observation consists of a set of probes, which
are deployed at specified locations. These probes interact with those
particles which pass through their locations, and thus measure their
properties. Realistically, these probes cannot distinguish between
separate particles, which means that, first, the cross-particle
velocity moments are difficult to capture, and, second, the
same-particle velocity moments are ensemble-averaged over those
particles which interact with a particular probe.

Let us assume that a probe is placed in the location $\BV y$ in the
$3$-dimensional physical place, and let us presume that $N_{\BV y}$
particles (out of total $N$) interact with that probe during the
measurement. Then, the ``energetic'' quantity, which can be somewhat
easily captured, is the ensemble average
\begin{equation}
\BV E(\BV y)=\frac 1{N_{\BV y}}\sum_{i=1}^{N_{\BV y}}\diag(\bE_i),
\end{equation}
where $\bE_i$ are the $3\times 3$ blocks of $\bE$ which correspond to
the quadratic velocity self-moments of particles passing through $\BV
y$, and lie on the main diagonal of $\bE$, whereas ``$\diag$'' denotes
the operation of extracting the main diagonal from $\bE_i$ and mapping
it into a vector. Thus $\BV E(\BV y)$ contains the average wind energy
at the location $\BV y$, measured in all three directions separately.
The total scalar kinetic energy of the wind is, obviously, the half of
the trace of $\BV E(\BV y)$.

Typically, multiple probes in different locations are used, such that
the distances between measured particles are known and given by the
coordinate offsets of the probes. The subsequent DFT over the set of
probe coordinates $\BV y$ reveals the scaling structure of these
quantities in the bundles $\cB_{ij}$ in \eqref{eq:B_ij}, as described
above in Section \ref{sec:structures}.  If, by $\BV k$, we denote the
Fourier wavevector of the DFT in $\BV y$, then the corresponding
Fourier transform $\BV E_{\BV k}$ will apparently have the same bulk
scaling structure as its counterpart $\bE_{\BV K}$ in the bundles
$\cB_{ij}$ (remember that the scaling is shared across the
bundles). Therefore, from the estimates in \eqref{eq:tf_2} and
\eqref{eq:el_2}, we arrive at the following relations:
\begin{enumerate}[\indent a.]
\item {\bf The Thomas--Fermi potential.} For the Thomas--Fermi
  potential, $\BV E_{\BV k}$ is estimated to scale as
\begin{equation}
\label{eq:tf}    
\BV E_{\BV k}\sim\|\BV k\|^{-5/3}
\end{equation}
at small inertial scales.  At large inertial scales, the Thomas--Fermi
potential is unlikely to have any discernible effect on dynamics.
\item {\bf The electrostatic potential.} For the electrostatic
  potential, $\BV E_{\BV k}$ is estimated to scale as
\begin{subequations}
\label{eq:el}
\begin{equation}
\BV E_{\BV k}\sim\|\BV k\|^{-3}\qquad\text{at large inertial scales},
\end{equation}
\begin{equation}
\BV E_{\BV k}\sim\|\BV k\|^{-5/3}\qquad\text{at small inertial
  scales}.
\end{equation}
\end{subequations}
\end{enumerate}
In laboratory experiments, the inverse five-thirds energy scaling at
near viscous ranges, as predicted above in \eqref{eq:tf} and
\eqref{eq:el} for both the Thomas--Fermi and the electrostatic
potentials, is observed rather reliably -- see, for example, the
recent work by Buchhave and Velte \cite{BucVel}. On the other hand, at
larger inertial scales, the results of observations sometimes do not
reveal any power scaling; for example, in \cite{BucVel}, the energy
spectrum flattens out at larger scales. This, however, does not mean
that the energy spectrum becomes constant, only that there is no
discernible linear trend on the log-log plot. According to our
hypothesis, such an uncertain behavior at large scales is to be
expected if the particles are indeed driven by the Thomas--Fermi or a
qualitatively similar (for example, the Ziegler--Biersack--Littmark
\cite{ZieBieLit}) short-range potential, whose effect does not extend
too far beyond the viscous scale.

\begin{figure}
\includegraphics[width=0.6\textwidth]{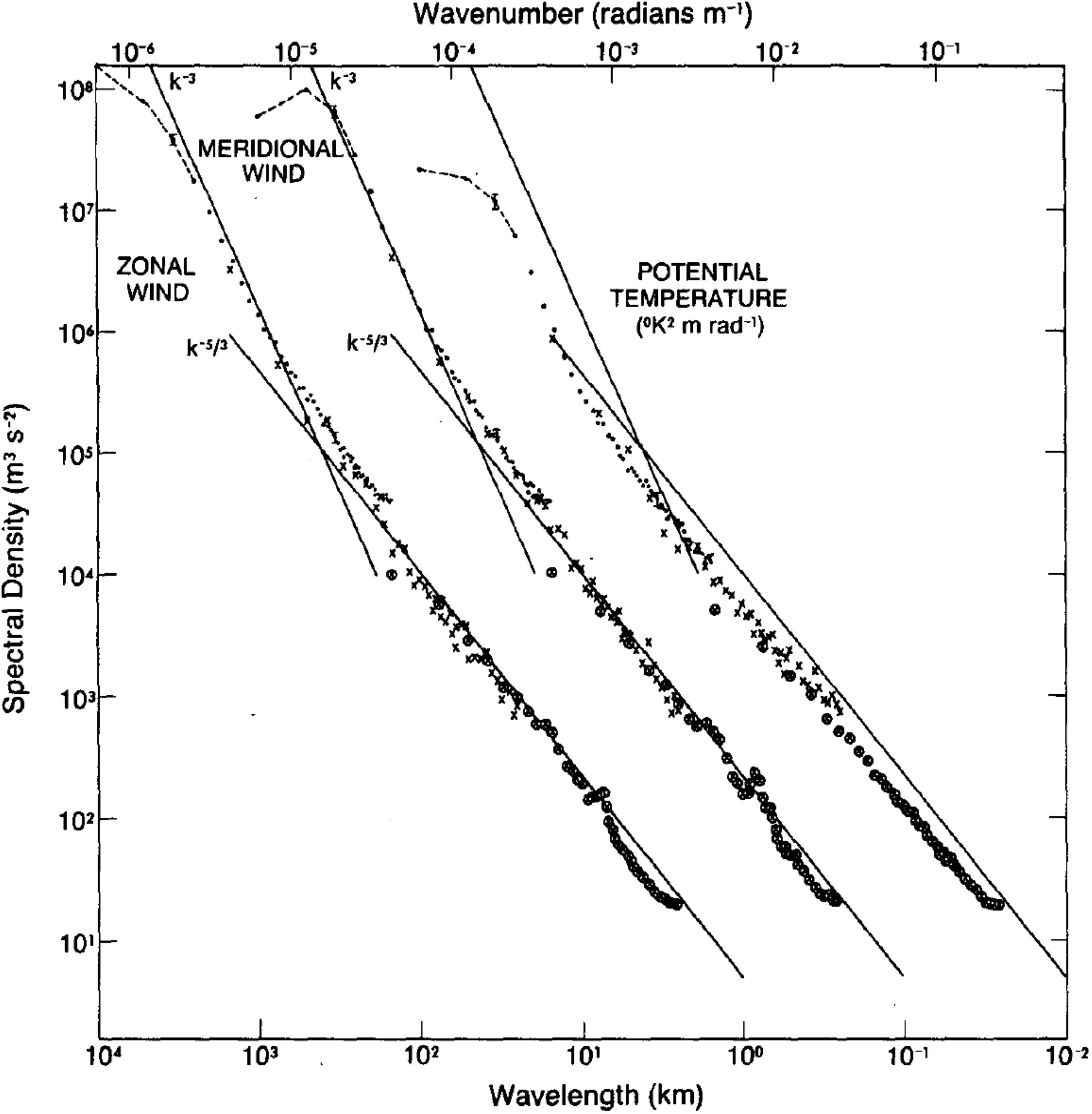}
\caption{Wind energy and temperature spectra, reproduced from Fig.~3
  in \cite{NasGag}. \copyright\,American Meteorological Society. Used
  with permission.}
\label{fig:nastrom_gage}
\end{figure}

Surprisingly, the observations of the Earth atmosphere capture a
radically different behavior, where the power scaling of the energy
spectrum is observed in a broad range of scales. A striking example is
the work of Nastrom and Gage \cite{NasGag}, where the observations
were obtained from the Global Atmospheric Sampling Program (GASP)
dataset. With the permission from the American Meteorological Society,
we reproduce Figure 3 from \cite{NasGag} in Figure
\ref{fig:nastrom_gage}, which shows the energy spectra of the
meridional (north--south) and zonal (east--west) winds. Both the
meridional and zonal wind energy spectra exhibit the inverse cubic
power scaling at large scales, and the inverse five-thirds power
scaling at small scales. According to Nastrom and Gage \cite{NasGag},
these trends appear to be universal, and largely independent on the
latitude, longitude and altitude of the flow.

The observations in \cite{NasGag} are not consistent with the
assumption that the motions at inertial scales are primarily driven
via the short-range Thomas--Fermi potential, as the latter lacks any
measurable ability to affect the scaling at larger scales. Yet, in
\cite{NasGag}, we see that the power scaling of the energy spectrum
manifests itself up to the synoptic scales, which exceeds the
effective range of the Thomas--Fermi potential by many orders of
magnitude. Also, somewhat paradoxically, the observed inverse cubic
spectrum at large scales and inverse five-thirds spectrum at small
scales match the predictions in \eqref{eq:el} for the electrostatic
interaction potential, which does not have a discernible effective
spatial range.

If the observed turbulent motions manifest according to our theory
above, the only rational explanation for the universality of the
spectra in \cite{NasGag} and in Figure \ref{fig:nastrom_gage} appears
to be that the wind energy spectra of the atmospheric turbulence are
indeed produced by the electrostatic potential. Recall that the Earth
atmosphere is not electrostatically neutral -- in fact, the density of
ionized molecules in the air is such that the average strength of the
electric field at low altitudes is about 130 volt per meter
\cite{AtmosElectricity}, so that its influence is not negligible.
While the relative density of the charged particles and ionized
molecules in the Earth atmosphere is relatively low (in comparison
with the electrostatically neutral molecules), a possible mechanism of
the statistical energy distribution is that the charged molecules
collide with the electrostatically neutral molecules around them and
``thermodynamically equilibrate'' their energy on a much shorter time
scale than the turbulent averaging time scale $T$.  While such a
hypothesis needs to be verified experimentally, it does nonetheless
offer a plausible explanation for the observed power scaling of the
atmospheric turbulent energy spectrum.

In addition to the meridional and zonal wind energy spectra, Nastrom
and Gage \cite{NasGag} examined the temperature spectrum, which is
also shown in Figure \ref{fig:nastrom_gage}. Observe that the
temperature spectrum exhibits the same scaling trends as the wind
energy spectra, however, presently we do not have a straightforward
explanation for this. The reason is that the temperature is,
effectively, the average kinetic energy of molecules, only measured in
the reference frame which moves with the average wind (and is
therefore subject to the turbulent fluctuations of the
wind). Therefore, the temperature is affected simultaneously by the
fluctuations of the kinetic energy of the wind, as well as the
fluctuations of the thermal motion in the wind reference frame, and it
is not clear which fluctuations provide the dominant contribution. In
addition, what is plotted in Figure \ref{fig:nastrom_gage} appears to
be the square of the temperature fluctuations, since the units are
indicated as ``$^\circ K^2$'' (that is, degrees Kelvin squared). Thus,
we presently refrain from further comments on the displayed
temperature spectrum, and will examine it in a separate work.

\section{Summary}

In the present work, we investigate the Liouville equation for $N$
particles, which interact via a generic potential. We find that a
strong large scale flow velocity creates forcing in the inertial
ranges of the energy equation via the potential coupling. This forcing
lives in the $3$-dimensional bundles of the full $3N$-dimensional
coordinate space, with each bundle belonging to an unordered pair of
particles. These bundles are destroyed in the course of the BBGKY
formalism, so that no such forcing manifests in the Boltzmann
equation, and, subsequently, in the Euler and Navier--Stokes
equations.  On the other hand, measurements can register the resulting
flow in these bundles -- effectively, these bundles are ``collapsed''
into the physical $3$-dimensional space, where the measurement probes
are located. Although an individual information in each bundle is lost
during such collapse, the bulk trends, shared across all bundles, may
persist.

Next, we investigate appropriately scaled time averages of solutions
of the Liouville equation. In particular, we develop crude estimates
for the power scaling of the energy spectrum in the inertial range,
assuming that it is driven by the strong large scale flow coupled to
the inertial scales via the interaction potential. The following is a
summary of assumptions and simplifications we utilize above to arrive
at the scaling estimates:
\begin{enumerate}[\indent 1.]
\item We assume that the spectrum of a physically relevant solution of
  the Liouville equation has three subranges: strong large scale flow,
  inertial range, and viscous range.
\item In the inertial range, we assume that the only measurable effect
  of the potential is the strong large scale velocity forcing in the
  energy equation.
\item We assume that the exponentially weighted time average via the
  Laplace transformation in Section \ref{sec:time_average} is a
  sufficiently close analog of a windowed time average.
\item For the time-averaged dynamics in Section \ref{sec:simplified},
  we make the following assumptions:
\begin{itemize}
\item On the chosen averaging time scale, the flow is effectively
  incompressible;
\item The centered skewness moment in the inertial range can be
  neglected in comparison with the velocity and energy moments;
\item As the inertial range transitions into the viscous range, the
  time-averaged velocity rapidly decays to zero, similarly to a
  ``cut-off''.
\end{itemize}
\item To make the spectrum decay estimates, we assume that the
  time-averaged solution is turbulent and its energy spectrum has
  power scaling, i.e.~the strong large scale flow velocity has an
  appropriate pattern to induce turbulence in the inertial range.
\end{enumerate}
Under these assumptions, we find that:
\begin{enumerate}[\indent a.]
\item In the near-viscous inertial subrange, the energy spectra are
  estimated to decay as the inverse five-thirds power of the Fourier
  wavenumber for the Thomas--Fermi interatomic potential
  \cite{Tho,Fer} (or a similar one, such as the
  Ziegler--Biersack--Littmark potential \cite{ZieBieLit}), as well as
  the electrostatic potential;
\item In the large scale inertial subrange, the energy spectra are
  estimated to decay as an inverse cubic power of the Fourier
  wavenumber for the electrostatic potential, while the Thomas--Fermi
  interatomic potential is not expected to affect the dynamics due to
  its short effective range.
\end{enumerate}
We compare the predictions with the measurements provided by Buchhave
and Velte \cite{BucVel} for the turbulent flow in laboratory
conditions, and by Nastrom and Gage \cite{NasGag} for the Earth
atmosphere. The measurements in \cite{BucVel} appear to be consistent
with the predictions for the Thomas--Fermi potential. Strikingly, the
observations of Nastrom and Gage \cite{NasGag} indicate that, at the
inertial scales, the Earth atmosphere behaves as if driven via the
electrostatic potential, i.e. exhibiting the inverse five-thirds power
spectrum at small scales, and the inverse cubic spectrum at large
scales. We suggest a hypothesis that, since the Earth atmosphere is
not electrostatically neutral, it could indeed be the case that the
atmospheric turbulent energy spectra are induced by the large scale
flow coupled via the electrostatic potential.

In laboratory conditions, this hypothesis can be tested experimentally
by measuring the turbulent energy spectra of a gas which is
electrostatically charged to varying degrees. For a fully neutral gas,
one should expect the spectra as in \cite{BucVel}, with the inverse
five-thirds scaling in near-viscous ranges due to the Thomas--Fermi
potential, and an indeterminate behavior at large scales. As the
electric charge of the gas increases, one should initially observe the
extension of the inverse five-thirds scaling onto larger scales (where
no discernible power scaling was observed for a neutral gas), and,
eventually, the transition to the inverse cubic power at large scales,
as in \cite{NasGag}.

If the atmospheric turbulent motions and the related energy spectra
are indeed caused by the electrostatic potential, it would be
interesting to examine the behavior of large, relatively dense systems
of celestial bodies (such as Saturnian rings, for example), which
interact via the gravitational potential. Recall that the
gravitational potential has the same form as the electrostatic
potential, except for the opposite sign -- while the electrostatically
charged particles repel, the bodies with mass attract. This means that
the same reasoning as above could likely be applied to the estimates
of the kinetic energy spectra of large systems of celestial bodies.

Finally, we have to point out that the conventional Euler and
Navier--Stokes equations of fluid dynamics are incapable of modeling
the described effects, because the bundles, in which the latter
manifest, are destroyed in the process of the BBGKY formalism. If it
is indeed confirmed that the turbulent motions and the related energy
spectra appear as described in the present work, the conventional
equations of the fluid dynamics will likely have to be appropriately
modified to extend their applicability onto turbulent motions in the
inertial ranges.

\appendix

\section{The structure of the potential}
\label{app:potential_structure}

Recalling \eqref{eq:Phi}, for the Fourier transform $\Phi_{\BV K}$ in
\eqref{eq:moments_fourier} we write
\begin{equation}
\Phi_{\BV K}=\int_{\RR^{3N}}\Phi(\BV X)e^{-\hi\BV K\cdot\BV X}\dif\BV
X=\sum_{i=1}^{N-1}\sum_{j=i+1}^N\int_{\RR^{3N}}e^{-\hi\BV K\cdot\BV X}
\phi\left(\|\BV x_i-\BV x_j\|\right)\dif\BV X.
\end{equation}
For each individual integral in the sum, we can write
\begin{multline}
\int_{\RR^{3N}}e^{-\hi\BV K\cdot\BV X}\phi\left(\|\BV x_i-\BV x_j\|
\right)\dif\BV X=\prod_{\myatop{m=1}{m\neq i,j}}^N\delta(\BV k_m)
\int_{\RR^6}e^{-\hi(\BV k_i\cdot\BV x_i+\BV k_j\cdot\BV x_j)}
\phi\left(\|\BV x_i-\BV x_j\|\right)\dif\BV x_i\dif\BV
x_j=\\=\delta(\BV k_i+\BV k_j)\prod_{\myatop{m=1}{m\neq
    i,j}}^N\delta(\BV k_m) \int_{\RR^3}e^{-\hi\BV k_i\cdot\BV
  y}\phi(\|\BV y\|)\dif\BV y.
\end{multline}
For the remaining $3D$-integral, we switch to the spherical coordinate
system $(r,\alpha,\beta)$, whose polar axis is aligned with $\BV k$,
such that $\BV k\cdot\BV y=\|\BV k\|r\cos\beta$. This yields
\begin{equation}
\int_{\RR^3}e^{-\hi\BV k\cdot\BV y}\phi(\|\BV y\|)\dif\BV
y=\int_0^\pi\int_0^{2\pi}\int_0^\infty e^{-\hi\|\BV k\|r\cos\beta}
\phi(r)r^2\sin\beta\dif r\dif\alpha\dif\beta.
\end{equation}
The integral over the angles alone yields
\begin{equation}
\int_0^\pi\int_0^{2\pi} e^{-\hi\|\BV k\|r\cos\beta} \sin\beta
\dif\alpha\dif\beta=\frac 2{\|\BV k\|r}\sin(2\pi\|\BV k\|r),
\end{equation}
which leads to
\begin{equation}
\int_{\RR^3}e^{-\hi\BV k\cdot\BV y}\phi(\|\BV y\|)\dif\BV y=\frac
2{\|\BV k\|}\int_0^\infty r\phi(r) \sin(2\pi\|\BV k\|r)\dif r.
\end{equation}
At this point, denoting $\phi_{\|\BV k\|}$ as in \eqref{eq:phik}, and
assembling the pieces together, we arrive at \eqref{eq:PhiK}.

\section{The averaging limit of the Laplace transformation}
\label{app:proof}

Here we follow the proof given in \cite[Section~2.1]{GluMil}. Let
$G(T)$ denote the $T$-window average of $F$:
\begin{equation}
G(T)=\frac 1T\int_0^T F(t)\dif t.
\end{equation}
Assuming that the limit of $G(T)$ as $T\to\infty$ exists, the Final
Value theorem states that
\begin{equation}
\lim_{T\to\infty} G(T)=\lim_{s\to 0} s\cL\{G\}(s).
\end{equation}
We are going to show that, in addition to the above identity,
\begin{equation}
\lim_{s\to 0} s\cL\{G\}(s)=\lim_{s\to 0} s\cL\{F\}(s).
\end{equation}
First, observe that
\begin{equation}
\cL\{G\}(s)=\int_0^\infty e^{-st}G(t)\dif t=\int_0^\infty e^{-st}
\left(\frac 1t\int_0^t F(\tau)\dif\tau\right)\dif t.
\end{equation}
At the same time, note that
\begin{equation}
\frac {e^{-st}}t=\int_s^\infty e^{-pt}\dif p,
\end{equation}
and, therefore,
\begin{multline}
\label{eq:a1}
\cL\{G\}(s)=\int_0^\infty\left(\int_s^\infty e^{-pt}\dif p\right)
\left(\int_0^t F(\tau)\dif\tau\right)\dif t=\int_0^\infty\dif t
\int_s^\infty\dif p\; e^{-pt} \int_0^t F(\tau)\dif\tau =\\=
\int_s^\infty\dif p\int_0^\infty\dif t\; e^{-pt} \int_0^t F(\tau)
\dif\tau =\int_s^\infty \cL\left\{\int_0^t F(\tau)
\dif\tau\right\}(p)\dif p.
\end{multline}
Above, Fubini's theorem was used to interchange the order of
integration in $\dif t$ and $\dif p$, given that all integrands are
nonnegative (as $F$ is a probability density), and under the further
assumption that the resulting integral is finite. To verify the
latter, observe that, from the derivative formula for the Laplace
transformation we know that
\begin{equation}
\label{eq:a2}
\cL\{F\}(p)=p\cL\left\{\int_0^t F(\tau)\dif\tau\right\}(p),\quad
\text{or}\quad\cL\left\{\int_0^t F(\tau)\dif\tau\right\}(p)=\frac
1p\cL\{F\}(p).
\end{equation}
Assuming that $F$ is bounded, its Laplace image is estimated as
\begin{equation}
\cL\{F\}(p)\sim\frac 1p,\quad\text{and, therefore,}\quad\frac
1p\cL\{F\}(p)\sim\frac 1{p^2}.
\end{equation}
Thus, the integral of the latter expression from $s$ to $\infty$ is
finite as long as $s>0$, which means that the use of Fubini's theorem
above is justified. Combining \eqref{eq:a1} and \eqref{eq:a2}, we
arrive at
\begin{equation}
\cL\{G\}(s)=\int_s^\infty\frac 1p\cL\{F\}(p)\dif p,\quad\text{or}\quad
s\cL\{G\}(s)=s\int_s^\infty\frac 1p\cL\{F\}(p)\dif p.
\end{equation}
The limit of the latter expression as $s\to 0$ can be computed via
L'H\^opital's ``$0/0$'' rule:
\begin{equation}
\label{eq:a3}
\lim_{s\to 0} s\cL\{G\}(s)=\lim_{s\to 0}\frac{\int_s^\infty\frac 1p
  \cL\{F\}(p)\dif p}{1/s}=\lim_{s\to 0} \frac{-\frac 1s\cL\{F\}(s)
}{-1/s^2}=\lim_{s\to 0}s\cL\{F\}(s).
\end{equation}
Finally, replacing $s=T^{-1}$ in the right-hand side and recalling
\eqref{eq:F_hat} yields \eqref{eq:laplace_average}.

\ack The author thanks the American Meteorological Society for
granting the permission to reproduce Figure 3 from the article by
Nastrom and Gage \cite{NasGag}. This work was supported by the Simons
Foundation grant \#636144.

\end{document}